 \documentclass[11pt]{article}

 \pdfoutput=1

 \usepackage[polutonikogreek,english]{babel}

 \usepackage{textgreek} % ex. \textgreek{>en'ergeia}

\usepackage{bigints}
\usepackage[round]{natbib} % omit 'round' option if you prefer square brackets

\usepackage{graphicx,siunitx,xspace}
\usepackage{amsmath}
\usepackage{amssymb}
\usepackage{hyperref}
\usepackage{enumitem}
\usepackage{color}
\usepackage{enumitem}
\usepackage{float}
\usepackage{amsfonts,bm}
\usepackage{epstopdf}
\usepackage[mathscr]{euscript}

  \usepackage[normalem]{ulem}
  \usepackage{textcomp}
\usepackage[utf8]{inputenc}
\usepackage[T1]{fontenc}
\usepackage{pifont} % \ding{number}

\usepackage{chemformula}

\usepackage{float}
\usepackage{ifthen}
\begin{document}
%=====================

%-------------------------------------------------------------------

%\title
\begin{center}
%- - - - - - - - - - - - - - - - - - - - - - - - 
{\bf \Large
An English translation of the paper of Max 
\citet[][]{Planck_absolute_Entropie_1916}}
%- - - - - - - - - - - - - - - - - - - - - - - - 
\\ \vspace*{2mm}
%- - - - - - - - - - - - - - - - - - - - - - - - 
\hspace*{-0mm}
{\bf \Large 
``\,\underline{Über die absolute Entropie einatomiger Körper}\,''}
%- - - - - - - - - - - - - - - - - - - - - - - - 
\\ \vspace*{2mm}
%- - - - - - - - - - - - - - - - - - - - - - - - 
{\bf \Large or: ``\,\underline{On the absolute entropy of monatomic bodies}\,''}
%- - - - - - - - - - - - - - - - - - - - - - - - 
\\ \vspace*{2mm}
%- - - - - - - - - - - - - - - - - - - - - - - - 
{\bf \Large
to provide a readable version of the German content.
}
%- - - - - - - - - - - - - - - - - - - - - - - - 
\\ \vspace*{2mm}
{\bf \large\color{blue}
Translated by Dr. Hab. Pascal Marquet 
}
%- - - - - - - - - - - - - - - - - - - - - - - - 
\\ \vspace*{2mm}
{\bf\bf\color{blue}  \large Possible contact at: 
    pascalmarquet@yahoo.com}
    \vspace*{1mm}
    \\
{\bf\bf\color{blue} 
    Web Google-sites:
    \url{https://sites.google.com/view/pascal-marquet}
    \\ ArXiv: 
    \url{https://arxiv.org/find/all/1/all:+AND+pascal+marquet/0/1/0/all/0/1}
    \\ Research-Gate:
    \url{https://www.researchgate.net/profile/Pascal-Marquet/research}
}
%- - - - - - - - - - - - - - - - - - - - - - - - 
\\ \vspace*{1mm}
\end{center}

\hspace*{65mm} Version-1 / \today

\vspace*{-2mm} 
\begin{center}
--------------------------------------------------- 
\end{center}
\vspace*{-11mm}

%%%%%%%%%%%%%%%%%%%%%%%%%%%%%%%%%%%%%%%%%%%%%%%%%%%%%%%%%%%%%%%%%%%%%
% REFERENCES
%%%%%%%%%%%%%%%%%%%%%%%%%%%%%%%%%%%%%%%%%%%%%%%%%%%%%%%%%%%%%%%%%%%%%
%{\small
\bibliographystyle{ametsoc2014}
\bibliography{Book_FAQ_Thetas_arXiv}
%}

\vspace*{-2mm} 
\begin{center}
--------------------------------------------------- 
\end{center}
\vspace*{-2mm}

Uncertainties/alternatives in the translation are indicated {\color{blue} (in blue)} with {\it\color{blue} italic terms}, together with some additional footnotes (indicated with {\it\color{blue} P. Marquet} or sometimes in {\it\color{magenta}magenta}). 
% Moreover, I have added some highlight (shown as \dashuline{\,dashing text}), in particular about the \dashuline{energy} concept.

I have written {\bf in bold} those parts of the text that deal in particular with the problem of determining {\bf the ``\,zero-point energy\,''} {\it\color{blue}(\,``\,Nullpunktsenergie\,''\,)}.
% in particular in the Section~\ref{Section-1} (R. Clausius, W. Thomson), in the Section~\ref{Section-2}, and at the end of the Section~\ref{Subsection-3-2}.
It is indeed for this aspect that I felt the need to translate the 2 thesis memoirs (1879 for the Doctor dissertation; 1880 for the Habilitation) and the 5 articles (1887 about the conservation of energy; 1915a,b,c and 1916 about the absolute definition of the entropy; 1943 about the discovery of quanta in physics), all written in German by Max Planck.

% Max Planck's 1887 contribution should be seen as nothing more than a synthesis of the late 1880s, with many concepts that would be strongly modified (or even invalidated) by the coming revolutions of relativistic and quantum mechanics (at the beginning of the 20th century).
I have, of course, kept \dashuline{the original text of Planck (in black)} unchanged, while sometimes including {\it\color{blue}\dashuline{additional notes (in blue)}}. % to put into perspective the concepts that are sometimes outdated (and considered invalid) today.

% Planck's text is organised almost linearly within each chapter, without any separation and most often with very long sentences, as is often the case in German. 
% I therefore wanted to add a number of section-like headings (in blue), in order to make Planck's chosen organisation clearer, with in the Section~\ref{Section-1} the names of the various scientists appearing just before the description of their contributions.

% In addition, as Max Planck refers many times to the pagination of his text for cross-references, I have added {\it\color{blue}\dashuline{(always in blue)}} the beginning of each page (more or less one sentence, depending on the pages).  

Do not hesitate to contact me in case of mistakes or any trouble in the English translation from the German text.

\vspace*{-4mm}
% ``{\it \/}''
\begin{center}
========================================================
\end{center}
\vspace*{-11mm}

%\newpage
%==========================
  \tableofcontents
%==========================

\newpage
\begin{center}
%==========================================================
\underline{\Large\bf On the absolute entropy of monatomic bodies}
\vspace*{2mm}

{\large\bf by Max Planck (8 June 1916)}
%==========================================================
\end{center}
\vspace*{-4mm}

%========================================
\section{\underline{Introduction} (p.653-655)}
%========================================
\label{Section-1-Introduction}
\vspace*{-2mm}
% Einleitung.
% Introduction

% I. Die Frage nach dem Werte der absoluten Entropie eines Körpers, im Sinne des Nernstschen Wärmetheorems, hängt aufs engste zusammen mit derjenigen nach der physikalischen Struktur des Phasenraums, welche durch die Größe, Form und Lage der Elementargebiete der Wahrscheinlichkeit bedingt wird. Denn sobald diese bekannt ist, läßt sich durch ein eindeutiges kombinatorisches Verfahren die thermodynamische, ganzzahlige Wahrscheinlichkeit $W$ und daraus die Entropie des Körpers $k-In W$ berechnen. 

§\:1. 
The question of the value of the absolute entropy of a body, in the sense of Nernst's heat theorem, is closely related to that of the physical structure of the phase space, which is determined by the size, shape and position of the elementary areas of probability. As soon as this is known, the thermodynamic, integer probability $W$, and from this the entropy of the body $k\:.\:\ln(W)$ can be calculated using a clear combinatorial 
%- - - - - - - - - - - -
procedure{\color{blue}$\,$\footnote{$\:${\it\color{blue}Note that in all next relationships, ``\,$k$\,'' is the Planck-Boltzmann's constant / P. Marquet}.}}. 
%- - - - - - - - - - - -

% Wenn z.B. der Phasenraum eine dreifach unendliche Anzahl von Elementargebieten enthält, so daß ein bestimmtes Elementargebiet durch 3 voneinander unabhängige Ordnungszahlen $n, n', n''$ charakterisiert wird, so ergibt sich für die thermodynamische charakteristische Funktion $\Psi$, d. h. den negativen Quotienten aus der freien Energie $F$ und der Temperatur $T$:

If, for example, the phase space contains a triple infinite number of elementary regions, so that a certain elementary region is characterized by $3$ independent atomic numbers $n, n', n''$, then the thermodynamic 
%- - - - - - - - - - - -
characteristic function results in $\Psi\,${\color{blue}\footnote{$\:${\it\color{blue}Note that the notation and the meaning of $\Psi$ correspond to the first of the two ``\,characteristic functions\,''
$\psi=S-U/T$ and $\psi'=S-(U+P\:V)/T$
introduced first by the French scientist Fran\c{c}ois Massieu in 1869, thus before that the (unnamed) functions 
$\psi=\epsilon-t\:\eta$ (for $F=U - T\:S$) and  
$\zeta=\epsilon+p\:v-t\:\eta$ (for $G=U +P\:V - T\:S$)
will be introduced by Gibbs (1875-78), and also before that the free-energy function 
$F=U - T \: S$ will be introduced (and named) by Helmholtz (1882-83). Max Planck has used and studied the Massieu's function  $\,\psi=-\,F/T$ in all his German papers and book, with the interest that they are modified entropy functions and, as such, are more suitable for direct studies of maximum entropy states, for instance / P. Marquet.}}},
%- - - - - - - - - - - -
i.e.  the negative quotient of the free energy $F$ and the 
%- - - - - - - - - - - -
temperature\,{\color{blue}\footnote{$\:${\it \color{blue}Note that from the two relationships $\Psi=-F/T$ and $F=-\:k\:T\:\ln(Z)$, where $Z$ is the partition function defined in statistical physics, the Massieu's characteristic function $\Psi$ defined by (\ref{label_eq_1}) simply corresponds to $\Psi=k\:\ln(Z)$, with therefore the partition function $Z$ made of the triple sum forming the argument of the logarithm in (\ref{label_eq_1}) / P. Marquet.}}} 
%- - - - - - - - - - - -
$T$:$\,${\color{blue}\footnote{$\:$Max Planck, Verhandl. d. Deutsch. Phys. Ges. 17, S. 444. 1915 
{\it \color{blue}(Namely:
``\,Die Quantenhypothese f\"ur Molekeln mit mehreren Freiheitsgraden 
 / The quantum hypothesis for molecules with multiple degrees of freedom\,'' 
(Erste und zweite Mitteilung / First and second communication). 
Berichte der Deutschen Physikalischen Gesellschaft (Reports of the German Physical Society), Vol.17, N°24, p.407-418 and p.438-451 / P. Marquet).}
%- - - - - - - - - - - -
}}
\vspace*{-3mm}
%- - - - - - - - - - - -
\begin{align}
\Psi & \: = \; -\:\frac{F}{T} \; = \; 
k \; \ln\left[
\;\;\sum_{n=0}^{\infty}
\;\;\sum_{n'=0}^{\infty}
\;\;\sum_{n''=0}^{\infty}
\;\;p_{\,n\:n'\:n''} \:\:.\:\:
\exp\left(-\:\frac{\overline{U}_{n\:n'\:n''}}{k\:T}\right)
\:\;\right]
\label{label_eq_1} \: .
\end{align}
%- - - - - - - - - - - -
From this the energy:$\,${\color{magenta}\footnote{\it\color{magenta}$\:$In fact, nowadays we know that  only the difference in internal energy $\:U-U_{ref}$ can be computed from (\ref{label_eq_2}) and the Massieu's function $\Psi=k\:\ln(Z)$ depending on the statistical-physics partition function $Z$. / P. Marquet}}
%- - - - - - - - - - - -
\vspace*{-3mm}
\begin{align}
U {\color{magenta}\;-\: U_{ref}} 
& \: = \; T^2\:.\:\,\frac{\partial\:\Psi}{\partial\:T}
\label{label_eq_2} \: 
\end{align}
%- - - - - - - - - - - -
and the entropy:$\,${\color{magenta}\footnote{\it\color{magenta}$\:$Like in (\ref{label_eq_2}) (a priori) only the difference in entropy $\:S-S_{ref}$ can be computed in (\ref{label_eq_3}) from the Massieu's function $\Psi=k\:\ln(Z)$ depending on the statistical-physics partition function $Z$. However, the Third-law hypothesis and the Planck's absolute definition of $S$ corresponds to $S_{ref}=0$ / P. Marquet}}
%- - - - - - - - - - - -
\vspace*{-3mm}
\begin{align}
S {\color{magenta}\;-\: S_{ref}} 
& \: = \; \Psi \;+\; T\:\frac{\partial\:\Psi}{\partial\:T}
\;=\; \Psi \;+\; \frac{U{\color{magenta}\;-\: U_{ref}} }{T}
\label{label_eq_3} \: .
\end{align}
% Hierbei bedeutet $U$ die mittlere Energie im Elementargebiet $(nn'n”)$, und die ganze Zahl $p$ das Verhältnis der Größe dieses Gebiets zu derjenigen des Elementargebiets (000). 
Here $\overline{U}_{n\:n'\:n''}$ means the average energy in the elementary region $(n\:n'\:n'')$, and the integer $p$ means the ratio of the size of this region to that of the elementary region $(000)$.

% Für die Bestimmung der Elementargebiete gelten folgende Regeln. Bezeichnen $\phi...$ die Lagenkoordinaten, $\psi...$ die dazugehörigen Impulskoordinaten des Phasenraums, so werden die Grenzflächen der Elementargebiete bestimmt durch die Gleichungen: $(4)$ wo $g g' g''$ gewisse für die Struktur des Phasenraumes charakteristische Funktionen der Phasenkoordinaten, und $n, n', n''$ voneinander unabhängige positive ganze Zahlen, einschließlich Null, bezeichnen. 

% The following rules apply to determining the elementary areas. If $\phi...$ denote the position coordinates, $\psi...$ the associated momentum coordinates of the phase space, then the boundary surfaces of the elementary regions are determined by the equations: $(4)$ where $g g' g''$ certain for the structure of the phase space denotes characteristic functions of the phase coordinates, and $n, n', n''$ denote independent positive integers, including zero.

The following rules apply to the determination of the elementary areas. If $\phi_1, \:\phi_2, \:.\:.\:. $ denotes the position coordinates, $\psi_1, \:\psi_2, \:.\:.\:. $ the corresponding %- - - - - - - - - - - - 
momentum coordinates$\,${\color{blue}\footnote{$\:$\it\color{blue}Note that the position coordinates $\psi_1, \:\psi_2, \:.\:.\:. $ do not correspond to the specific (atomic) characteristic function noted $\psi=\Psi/N$ by Max Planck and introduced in the following / P. Marquet).}}
%- - - - - - - - - - - - 
of the phase space, then the boundary surfaces of the elementary regions are determined by the equations 
\begin{align}
        g & \: = \; n  \; h \; , 
\quad   g'  \; = \; n' \; h \; , 
\quad   g'' \; = \; n''\; h \; , 
\label{label_eq_4}
\end{align}
where $g,\: g',\: g''$ denote certain functions of the phase coordinates which are characteristic of the structure of the phase space, and $n, \: n', \: n''$ denote mutually independent positive integers, including zero. 

% Für ein Differentialgebiet des Phasenraumes gilt dann die Beziehung:
  The relationship then applies to a differential region of the phase space:
% For a differential area of the phase space, the relationship then applies:
\begin{align}
 dG & \:\: = \; 
\bigintsss\!\!\!
\bigintsss\!\!\!
\bigintsss\!\!\!
\bigintsss\!\!\!
\bigintsss_{\;\;\;\;\;\;\;g,\:\;\;\;\;\;\;\;g',\:\;\;\;\;\;\;\;\;g''}^{\,g+dg,\:g'+dg',\:g''+dg''}
\!\!\!\!\!\!\!\!
.\,.\,.\;\: d\phi_1 \; d\phi_2 \:\:
.\,.\,.\:\: d\psi_1 \; d\psi_2 \:\:
.\,.\,.
\;=\;
d\left(    g^{\,f}\right) .\;
d\left(  g'^{\,f'}\right) .\;
d\left(g''^{\,f''}\right) 
\label{label_eq_5} \; .
\end{align}
% Hier ist die Integration über alle Phasenpunkte zu erstrecken, welche innerhalb der angegebenen Grenzen liegen. $f+f'+f''$ ist die Anzahl der Freiheitsgrade des Systems, von denen $f$, $f'$ und $f''$ miteinander kohärent sind.
% Here the integration must extend over all phase points that lie within the specified limits. $f+f'+f''$ is the number of degrees of freedom of the system, of which $f$, $f'$ and $f''$ are coherent with each other.
 Here the integration is to be extended over all phase points which lie within the specified limits, and $f+f'+f''$ is the number of degrees of freedom of the system, of which $f$, $f'$ and $f''$ are coherent with each other.
 
% Diesen Festsetzungen gemäß besitzt das Elementargebiet $(n n' n'')$, definiert durch die Grenzen $gn...$, die Größe: 
According to these determinations, the elementary region $(n \: n' \: n'')$, defined by the boundaries 
$g_{\,n}$, $\:g_{\,n+1}$, 
$\:g'_{\,n'}$, $\:g'_{\,n'+1}$,
$\:g''_{\,n''}$, $\:g''_{\,n''+1}$, 
has the size:
% According to these definitions, the elementary area $(n n' n'')$, defined by the boundaries $gn...$, has the size: 
\begin{align}
\bigintsss_{\,n}^{\,n+1}
\bigintsss_{\,n'}^{\,n'+1}
\bigintsss_{\,n''}^{\,n''+1}
 dG 
 & \:\: = \; 
\left( g_{\,n+1}^{\,f} \:-\: g_{\,n}^{\,f} \right) 
\:.\:
\left({g'}_{n'+1}^{\,f'}\:-\:{g'}_{n'}^{\,f'}\right) 
\:.\:
\left({g''}_{\!n''+1}^{\,f''}\:-\:{g''}_{\!n''}^{\,f''}\right) 
\label{label_eq_6} \; .
\end{align}
% Daher ist die Größe des Elementargebiets (000): $h...$, und die Werte p und U ergeben sich als: 
%Therefore the size of the elementary region (000) is: $h...$, and the values p and U are given as: $(7)(8)$, welche Ausdrücke, in (1) eingesetzt, den Wert von $\Psi$ und alles andere liefern.
%- - - - - - - - - - - -
Therefore{\color{blue}$\,$\footnote{$\:${\it\color{blue}In the next relationships, ``\,$h$\,'' is the Planck's constant minimal quantum  value of action / P. Marquet}.}}, 
%- - - - - - - - - - - -
the size of the elementary area $(000)$ is $h^{\,f+f'+f''}$, and the values $p$ and $\overline{U}$ are obtained as: 
\begin{align}
p_{\,n\:n'\:n''}
 & \:\: = \; 
 \left[\, \left( n\,+\,1 \right)^{\,f} 
    \:-\: \left( n \right)^{\,f} 
 \,\right]
 \;.\;
 \left[\, \left( {n'}\,+\,1 \right)^{\,f'} 
    \:-\: \left( {n'} \right)^{\,f'} 
 \,\right]
 \;.\;
 \left[\, \left( {n''}\,+\,1 \right)^{\,f''} 
    \:-\: \left( {n''} \right)^{\,f''} 
 \,\right]
\label{label_eq_7} \; , \\
\overline{U}_{\,n\:n'\:n''}
 & \:\: = \; 
\frac{\displaystyle
\bigintsss_{\,n}^{\,n+1}
\bigintsss_{\,n'}^{\,n'+1}
\bigintsss_{\,n''}^{\,n''+1}
U \:\:
d\left(    g^{\,f}\right) 
d\left(  g'^{\,f'}\right) 
d\left(g''^{\,f''}\right) 
}{p\;.\;h^{\,f+f'+f''}} 
\label{label_eq_8} \; ,
\end{align}
% which expressions, inserted into (1), give the value of $\ Psi$ and everything else deliver.
which expressions, inserted in (\ref{label_eq_1}), give the value of $\Psi$ and everything else.

% Dieser ganzen Berechnung liegt die physikalische Hypothese zugrunde, daß innerhalb eines jeden Elementargebietes die Phasenpunkte beliebige Lagen haben können --eine Annahme, die ich ausdrücklich als noch nicht zweifelsfrei hinstellen möchte, namentlich, da ihr gegenüber die andere Annahme, daß die Phasenpunkte nur an den Grenzen der Elementargebiete liegen können, gewisse Vorzüge zu haben scheint. 
% This entire calculation is based on the physical hypothesis that the phase points can have any position within each elementary region - an assumption that I would like to expressly state as not yet beyond doubt, especially since it is opposed to the other assumption that the phase points are only at the The boundaries of the elementary areas seem to have certain advantages.
 This whole calculation is based on the physical hypothesis that within each elementary region the phase points can have arbitrary positions (an assumption which I would like to state explicitly as not yet beyond doubt), especially since the other assumption (that the phase points can only lie at the boundaries of the elementary regions) seems to have certain advantages over it. 
    % p.655
% Indessen spielt für die vorliegende Untersuchung dieser Gegensatz, wie wir sehen werden, nur eine verhältnismäßig untergeordnete Rolle.
However, as we will see, this contrast plays only a relatively minor role in the present investigation.

% Als vorbereitendes Beispiel soll zunächst die Entropie eines einzelnen Atoms bestimmt werden. 
As a preliminary example, the entropy of a single atom should first be determined.

% \bigintssss \bigintsss \bigintss \bigints \bigint 
% \dashuline{\,dashing}  -{\it\color{blue}(change)}
%- - - - - - - - - - - -
% $\,$\footnote{$\:$.}  (\ref{label_eq_thermal_1})
% $\,$\footnote{$\:${\it\color{blue}()}.}

\vspace*{-2mm}
\begin{center}
------------------------------------------------------------------------------------
\end{center}
\vspace*{-8mm}

%========================================
\section{\underline{First part. {\color{blue}\it(Entropy for)} A single atom.} (p.655-659)}
%========================================
\label{Section-2-single-atom}
\vspace*{0mm}
% Erster Teil. Ein einzelnes Atom. 
% First part. A single atom.

% \bigintssss \bigintsss \bigintss \bigints \bigint 
% \dashuline{\,dashing}  -{\it\color{blue}(change)}
%- - - - - - - - - - - -
% $\,$\footnote{$\:$.}  (\ref{label_eq_thermal_1})
% $\,$\footnote{$\:${\it\color{blue}()}.}

%========================================
\subsection{\underline{Atom in a rectangular parallelepiped.} (p.655-657)}
%========================================
\label{SubSection-2-1-single-atom-rect-box}
\vspace*{0mm}

% §\:2. 2. Atom in einem rechtwinkligen Parallelepiped. 
% Atom in a rectangular parallelepiped.

\noindent  §\:2. 
% Wenn ein punktförmiges Atom in einem hohlen rechtwinkligen Parallelepiped von starr-elastischer Wandung mit den Kantenlängen $a, b, c$ frei herumfliegt, so bleiben die 3 Komponenten seiner Geschwindigkeit an Größe konstant, vertauschen aber in regelmäßig wiederkehrenden Intervallen ihr Vorzeichen in das entgegengesetzte. Bezeichnet m die Masse des Atoms, $AAAA$ die Anfangswerte ($t = 0$) der Koordinaten, $BBBB$ die der dazugehörigen Impulskoordinaten, so wird die Be- wegung des Atoms für alle Zeiten $t$ dargestellt dureh die folgenden 6 Gleichungen für die Koordinaten des entsprechenden Phasenpunktes in dem sechsdimensionalen Phasenraum: $(9)-(10)$, wobei zur Abkürzung gesetzt ist : $$AA$$. 
% If a point-shaped atom flies freely around in a hollow rectangular parallelepiped with a rigid-elastic wall with edge lengths $a, b, c$, the three components of its speed remain constant in size, but swap their sign to the opposite at regularly recurring intervals. If m denotes the mass of the atom, $AAAA$ the initial values ($t = 0$) of the coordinates, $BBBB$ those of the associated momentum coordinates, then the movement of the atom for all times $t$ is represented by the following 6 equations for the coordinates of the corresponding phase point in the six-dimensional phase space: $(9)-(10)$, where the abbreviation is: $$AA$$.
% - - - - - - - - - - - - - - - - - - - - - - - - - - - - -
If a point-shaped atom flies around freely in a hollow {\it\color{blue}(empty)} rectangular parallelepiped of rigid-elastic wall with the edge lengths $(a,\:b,\:c)$, the three components of its velocity remain constant in magnitude, but swap their sign to the opposite at regularly recurring intervals. If $m$ denotes the mass of the atom, $(x_0,\:y_0,\:z_0)$ the initial values ($t = 0$) of the coordinates, $(\xi_0,\:\eta_0,\:\zeta_0)$ those of the corresponding momentum coordinates, the movement of the atom for all times $t$ is represented by the following $6$ equations for the coordinates of the corresponding phase point in the six-dimensional phase space: 
%====================================
\begin{equation}
\left.
\begin{aligned}
%-------------------------------------------------------
  x  & \: = \; \frac{a}{2}
    \; - \; \frac{4\:a}{\pi^2} 
  \left[\;
   \cos\left(\alpha\right)
    \;+\;
   \frac{\cos\left(3\:\alpha\right)}{9}
    \;+\;
   \frac{\cos\left(5\:\alpha\right)}{25}
    \;+\;
   \,.\,.\,.\,
  \:\right]  
   \;  \\
%-------------------------------------------------------
  y  & \: = \; \frac{b}{2}
    \; - \; \frac{4\:b}{\pi^2} \,
  \left[\;
   \cos\left(\beta\right)
    \;+\;
   \frac{\cos\left(3\:\beta\right)}{9}
    \;+\;
   \frac{\cos\left(5\:\beta\right)}{25}
    \;+\;
   \,.\,.\,.\,
  \:\right]  
   \;  \\
%-------------------------------------------------------
  z  & \: = \; \frac{c}{2}
    \; - \; \frac{4\:c}{\pi^2} \,
  \left[\;
   \cos\left(\gamma\right)
    \;+\;
   \frac{\cos\left(3\:\gamma\right)}{9}
    \;+\;
   \frac{\cos\left(5\:\gamma\right)}{25}
    \, \;+\;
   \,.\,.\,.\,
  \:\right]  
   \;  
%-------------------------------------------------------
\end{aligned}
\;\;\;\;
\right\} 
\label{label_eq_9} \; ,
\end{equation}
%====================================
\vspace*{-4mm}
\begin{align}
\xi
 & \:\: = \; m \: \frac{dx}{dt} \;=\; \pm\: \xi_0 \; , 
 \quad\quad
\eta \; = \; m \: \frac{dy}{dt} \;=\; \pm\: \eta_0 \; ,  
 \quad\quad
\zeta \; = \; m \: \frac{dz}{dt} \;=\; \pm\: \zeta_0 \; ,  
\label{label_eq_10} 
\end{align}
where are set as abbreviations: 
\begin{align}
\alpha & \: = \; \frac{\pi}{a} \: 
\left( \frac{\xi_0 \; t}{m}    \:+\: x_0 \right) \; ,
    \quad\quad
\beta   \; = \; \frac{\pi}{b} \: 
\left( \frac{\eta_0 \; t}{m}   \:+\: y_0 \right) \; ,
    \quad\quad
\gamma   \; = \; \frac{\pi}{c} \: 
\left( \frac{\zeta_0 \; t}{m}  \:+\: z_0 \right) \; .
\nonumber % \label{label_eq_10} 
\end{align}

% Für die Quantenteilung des Phasenraums ist vor allem der Satz maßgebend, daß eine jede Phasenbahn ihrer ganzen Länge nach innerhalb eines und desselben Elementargebiets verläuft. Da nun die Raumkoordinaten $x, y, z$ des Atoms, wie man sich leicht nach den Gleichungen (9) überzeugen kann, im Laufe der Zeit denen eines jeden beliebigen Punktes innerhalb des Parallelepipeds beliebig nahe kommen, abgerechnet gewisse rationale Fälle, die für die Allgemeinheit nicht von Belang sind, so enthält jedes Elementargebiet des Phasenraumes sämtliche Punkte des Parallelepipeds, dagegen nur eine beschränkte Anzahl von Punkten des Gebiets der Impulskoordinaten $DDD$, und zwar zerfällt $dG$ gemäß (5) in 3 voneinander völligunabhängige Faktoren: $$AA$$.
%- - - - - - - - - - - - - - -
% The decisive factor for the quantum division of phase space is the principle that every phase path runs along its entire length within one and the same elementary region. Since the spatial coordinates $x, y, z$ of the atom, as one can easily see from equations (9), over time come arbitrarily close to those of any point within the parallelepiped, excluding certain rational cases for the generality is not important, then each elementary region of the phase space contains all the points of the parallelepiped, but only a limited number of points of the a of the momentum coordinates $DDD$, namely $dG$ breaks down into 3 completely independent factors according to (5): $ $AA$$.
%- - - - - - - - - - - - - - -
For the quantum division of phase space, the theorem that each phase trajectory runs along its entire length within one and the same elementary region is decisive. Since the spatial coordinates $(x,\:y,\:z)$ of the atom, as can easily be seen from equations (\ref{label_eq_9}), come arbitrarily close to those of any point within the parallelepiped over time (except for certain rational cases which are not relevant to the generality), then each elementary region of the phase space contains all points of the parallelepiped, but only a limited number of points of the region of momentum coordinates $(\xi,\:\eta,\:\zeta)$, and $dG$ decomposes according to (\ref{label_eq_5}) into $3$ completely independent factors:
\begin{align}
 dG & \:\: = \; 
\bigintsss\!\!\!
\bigintsss\!\!\!
\bigintsss\!\!\!
\bigintsss\!\!\!
\bigintsss\!\!\!
\bigintsss
dx\:.\:dy\:.\:dz\:.\:d\xi\:.\:d\eta\:.\:d\zeta
\;=\;
dg\:.\:dg'\,.\:dg'' 
\;\; . \nonumber  % \label{label_eq_5} \; .
\end{align}
% Die 3 Freiheitsgrade sind also inkohärent:  $$AA$$, und  $$BB$$, folglich, durch Ausführung der Integrationen, mit Rücksicht darauf, daß Jedes Elementargebiet ebensoviel positive wie negative Werte seiner Impulskoordinaten umfaßt: $$(11)$$.
% The 3 degrees of freedom are therefore incoherent: $$AA$$, and $$BB$$, consequently, by carrying out the integrations, taking into account that each elementary region includes as many positive as negative values of its momentum coordinates: $$(11)$$ .
The $3$ degrees of freedom are therefore incoherent, 
{\it\color{blue}with accordingly} 
\begin{align}
   f & \: = \; 1 \; , 
\quad\quad
   f'  \; = \; 1 \; , 
\quad\quad
   f'' \; = \; 1 \; , 
\nonumber % \label{label_eq_4}
\end{align}
and 
\vspace*{-3mm}
\begin{align}
   d\,g   \: = \bigintsss d\,x \:.\: d\,\xi \; , 
\quad\quad
   d\,g'  \: = \bigintsss d\,y \:.\: d\,\eta \; , 
\quad\quad
   d\,g'' \: = \bigintsss d\,z \:.\: d\,\zeta \; . 
\nonumber % \label{label_eq_4}
\end{align}
Therefore, by performing the integrations, taking into account that each elementary domain comprises as many positive as negative values of its momentum coordinates: 
\begin{align}
   g & \: = \; a\:.\:2\;\xi \; , 
\quad\quad
   g'  \; = \; b\:.\:2\;\eta \; , 
\quad\quad
   g'' \; = \; c\:.\:2\;\zeta \; . 
\label{label_eq_11}
\end{align}
% Aus diesen Werten folgt nach (7): $p = 1$, und nach (8), da die Energie des Atoms:$$AA$$, also durch Integration, mit Rücksicht auf (4):  $$BB$$. 
% From these values follows according to (\ref{label_eq_7}) $p = 1$, and according to (\ref{label_eq_8}), since the energy of the atom: since the energy of the atom: $$AA$$, thus by integration, with respect to (4): $$BB$$.
From these values it follows from (\ref{label_eq_7}) $p = 1$, and from (\ref{label_eq_8}) with the energy of the atom  
\begin{align}
 u & \:\: = \; 
 \frac{1}{2\:m} \:
 \left(\:
  \xi^2 \:+\: \eta^2 \:+\:\zeta^2 
 \:\right)
 \;=\;
 \frac{1}{8\:m} \:
  \left(\:
  \frac{g^2}{a^2} \:+\: 
  \frac{{g'}^{\,2}}{b^2} \:+\: 
  \frac{{g''}^{\,2}}{c^2} 
 \:\right)
 \nonumber % \label{label_eq_6}
\; , \\
 \overline{u}_{\,n\:n'\:n''}
 & \:\: = \; 
 \frac{1}{h^3}\:\:.
\bigintsss_{\,n}^{\,n+1}
\bigintsss_{\,n'}^{\,n'+1}
\bigintsss_{\,n''}^{\,n''+1} u\:.\:dg\:.\:dg'\,.\:dg'' 
 \nonumber \; ,  % \label{label_eq_6} \; .
\end{align}
by integration and taking into account (\ref{label_eq_4}): 
\begin{align}
 \overline{u}_{\,n\:n'\:n''}
 & \:\: = \; 
 \frac{h^2}{8\:m}\:
 \left(\:
  \frac{\:{n}^2\:+\:n\:+\:1/3\:}{a^2} 
  \;+\;
  \frac{\:{n'}^{\,2}\:+\:n'\:+\:1/3\:}{b^2} 
  \;+\;
  \frac{\:{n''}^{\,2}\:+\:n''\:+\:1/3\:}{c^2} 
 \:\right)
\label{label_eq_12} \; .
\end{align}
% Für die thermodynamische charakteristische Funktion des Atoms folgt daraus nach (1): $$(13)$$
For the thermodynamic characteristic function of the atom it follows from (\ref{label_eq_1}): 
\begin{align}
 \psi & \:\: = \; 
 k \; \ln\left[
\;\;\sum_{n=0}^{\infty}
\;\;\sum_{n'=0}^{\infty}
\;\;\sum_{n''=0}^{\infty} 
\;
\exp\left(-\:\frac{\overline{u}_{\,n\:n'\:n''}}{k\:T}\right)
\:\;\right]
\label{label_eq_13} \; .
\end{align}
% Am übersichtlichsten sind die beiden Grenzfälle hoher und tiefer Temperaturen. 
The two borderline cases of high and low temperatures are the clearest. 

% Bei hohen Temperaturen liefern nur die Summenglieder mit großen Ordnungszahlen $n, n', n''$ merkliche Beiträge zum Wert der Summe. Dann lassen sich die Summen ersetzen durch Integrale, und es wird $$AA$$, wo für $u$ der Ausdruck (12) einzusetzen ist. 
\dashuline{At high temperatures}, only the sum terms with large atomic numbers $(n, n', n'')$ make noticeable contributions to the value of the sum. Then the sums can be replaced by integrals and it becomes 
\begin{align}
 \psi & \:\: = \; 
 k \; \ln\left[\;
\bigintsss_{\,0}^{\infty}\!\!\!
\bigintsss_{\,0}^{\infty}\!\!\!
\bigintsss_{\,0}^{\infty}
\exp\left(-\:\frac{u}{k\:T}\right)
\:\:dn\:\:dn'\,\:dn'' 
\:\;\right]
 \nonumber \; ,  % \label{label_eq_6} \; .
\end{align}
where the expression (\ref{label_eq_12}) has to be inserted for $u$.

% Dies ergibt, mit Benutzung der Umformung: $$AA$$ und mit Weglassung verschwindend kleiner Glieder: $$(14)$$ 
This gives, using the transformation 
\vspace*{-1mm}
\begin{align}
 {n}^2 \:+\: n \:+\: \frac{1}{3}
 & \:\: = \; 
 \left(\:
  n \;+\;\frac{1}{2}
 \:\right)^{2}
 \:+\: \frac{1}{12}
\nonumber \;\: , % \label{label_eq_12} \; .
\end{align}
and omitting infinitesimally small terms: 
\vspace*{1mm}
\begin{align}
 \psi & \:\: = \; 
 k \; \left\{\;
\frac{3}{2}\;
\ln\left(\:
\frac{2\:\pi\:m\:k\:T}{h^2}
\:\right)
\;+\;
\ln\left(\:a\:b\:c\:\right)
\:\;\right\} 
 \label{label_eq_14} \; .
\end{align}
\vspace*{2mm}
% Daraus nach (2) die Energie: $$U$$ und nach (3) die Entropie: $$S(15)$$, von dem Ausdruck $\Psi$ nur unterschieden durch das Glied mit $e$. 
From this and according to (\ref{label_eq_2}), the energy {\color{blue}$u=T^2\:\partial\psi/\partial T$} is
\vspace*{-1mm}
\begin{align}
\boxed{\:
 u \; = \; \frac{3}{2} \: k \: T
 \:}
 \nonumber %  \label{label_eq_14}
 \; , 
\end{align}
and according to (\ref{label_eq_3}){\it\color{blue}, and thus 
$s=\psi+u/T=\psi+(3/2)\:k
=\psi+k\;(3/2)\:\ln(e)$ with $e=\exp(1)$,}
the entropy
\vspace*{-3mm}
\begin{align}
\boxed{\:
 s \; = \; 
 k \; \left\{\;
\frac{3}{2}\;
\ln\left(\:
\frac{2\:\pi\:e\:m\:k\:T}{h^2}
\:\right)
\;+\;
\ln\left(\:a\:b\:c\:\right)
\:\;\right\} 
\:}
\label{label_eq_15} \; ,
\end{align} 
only distinguished from the expression $\psi$ {\it\color{blue} given by (\ref{label_eq_14})} by the term with 
$e{\color{blue}\:=\exp(1)}$.

% Bei tiefen Temperaturen dagegen beschränkt sich die Summe (13) auf das erste Glied, also $$(16)$$
\dashuline{At low temperatures}, however, the sum (\ref{label_eq_13}) is limited to the first term, i.e. {\it\color{blue} from (\ref{label_eq_12}) with $n=0$\,}:
\begin{align}
 \psi {\:\color{blue}(T \approx 0)}
 \;=\; -\:\frac{1}{T} \:.\: \overline{u}_{\,0\:0\:0}
 &  \:\:=\; -\:\frac{1}{T} \:.\: 
  \frac{h^2}{24\:m}\:
 \left(\:
  \frac{1}{a^2} 
  \;+\;
  \frac{1}{b^2} 
  \;+\;
  \frac{1}{c^2} 
 \:\right)
 \label{label_eq_16} % \nonumber 
 \; .
\end{align}
% Daraus die Entropie s = 0 und die Nullpunktsenergie, entweder nach (2) oder direkt aus (12): $$(17)$$
From this and from 
{\bf \dashuline{the entropy $s = 0$}\,}, 
either according to (\ref{label_eq_2}) or directly from (\ref{label_eq_12}), 
{\bf \dashuline{the zero point energy}} 
%- - - - - - - - - - - - - - - - -
is:$\,${\color{blue}\footnote{$\:${\it\color{blue}This hypothesis ``\,$s = 0$ at $T \approx 0$~K\,'' is similar to the third-law of thermodynamics, which is valid for the more stable solid states, but here applied to a gas. 
If the zero point entropy $\overline{s}_{\,0\:0\:0}$ does not cancels out at $0$~K for a gas, then (\ref{label_eq_16}) should be replaced by: 
$\psi\,(T \approx 0) = \overline{s}_{\,0\:0\:0} 
- \overline{u}_{\,0\:0\:0}\,/\,T $,
and (\ref{label_eq_17}) by  
$u\,(T \approx 0) = \overline{u}_{\,0\:0\:0}
- T \:.\: \overline{s}_{\,0\:0\:0}$,
with $\:\overline{s}_{\,0\:0\:0}$ to be determined.
/ P. Marquet}.}}
%- - - - - - - - - - - - - - - - -
\vspace*{-1mm}
\begin{align}
 \boxed{\:
 u {\:\color{blue}(T \approx 0)} \;=\; \overline{u}_{\,0\:0\:0}
   \;=\;  
  \frac{h^2}{24\:m}\:
 \left(\:
  \frac{1}{a^2} 
  \;+\;
  \frac{1}{b^2} 
  \;+\;
  \frac{1}{c^2} 
 \:\right)
 \:}
 \label{label_eq_17} % \nonumber 
 \; .
\end{align}

% \bigintssss \bigintsss \bigintss \bigints \bigint 
% \dashuline{\,dashing}  -{\it\color{blue}(change)}
%- - - - - - - - - - - -
% $\,$\footnote{$\:$.}  (\ref{label_eq_thermal_1})
% $\,$\footnote{$\:${\it\color{blue}()}.}

\vspace*{-4mm}
\begin{center}
---------------------------------------------------
\end{center}
\vspace*{-3mm}

%========================================
\subsection{\underline{Atom in a hollow {\it\color{blue}(empty)} sphere} (p.657-659)}
%========================================
\label{SubSection-2-2-single-atom-sphere}
\vspace*{2mm}

% $ 3. Atom in einer Hohlkugel. 
% $ 3rd atom in a hollow sphere. 

\noindent §\:3. 
% Wenn ein Atom in einer Hohlkugel vom Radius $R$ mit starrelastischer Wandung frei herumfliegt, so bleibt es dauernd in der Ebene eines größten Kreises und beschreibt mit konstanter Geschwindigkeit y lauter gleichlange Strecken, deren Richtungen beim Auftreffen auf die Wandung mit dieser stets den nämliehen Winkel $\alpha$ bilden. Für $\alpha=0$ bewegt sich das Atom unmittelbar längs der Wandung, für $\alpha=\pi/2$ fliegt es durch den Mittelpunkt hin und her. 
If an atom flies around freely in a hollow {\it\color{blue}(empty)} sphere of radius $R$ with a rigid-elastic wall, it remains permanently in the plane of a largest circle and describes distances of equal length with a constant velocity $q$, the directions of which always form the same angle when hitting the wall form $\alpha$. For $\alpha=0$ the atom moves directly along the wall, for $\alpha=\pi/2$ it flies back and forth through the center.

% Von den 3 Freiheitsgraden dieses Systems sind 2 miteinander kohärent, da die Lage der Bahnebene gar keinen Einfluß auf die Quantenteilung des Phasenraums hat, also ist $ff$, und nach (5):   $dG$$, nach (7): $p=$, nach (8): $$(18)$$, und nach (1): $$(19)$$.
% - - - - - - - - - - - - - - - - -
Of the $3$ degrees of freedom of this system, $2$ are coherent with each other, since the position of the orbital plane has no influence at all on the quantum division of the phase space, so 
\vspace*{0mm}
\begin{align}
 f \; = \; 1 \; , 
 \quad\quad 
 f' \; = \; 2 \; , 
 \nonumber %  \label{label_eq_14}
\end{align}
and from (\ref{label_eq_5}): 
\vspace*{0mm}
\begin{align}
 d\,G \; = \; d\,g \:.\: d\,g'^{\,2} \; , 
 \nonumber %  \label{label_eq_14}
\end{align}
and from (\ref{label_eq_7}): 
\vspace*{0mm}
\begin{align}
 p \; = \; 2\:n' \:+\: 1 \; , 
 \nonumber %  \label{label_eq_14}
\end{align}
and from (\ref{label_eq_8}): 
\begin{align}
\overline{u}_{\,n\:n'}
 & \:\: = \; 
\frac{\displaystyle
\bigintsss_{\,n}^{\,n+1}
\bigintsss_{\,n'}^{\,n'+1}
u \:.\:
d\,g \:.\:
d\,g'^{\,2} 
}{( 2\:n' \:+\: 1 ) \;.\;h^{\,3}} 
\label{label_eq_18} \; ,
\end{align}
and from (\ref{label_eq_1}): 
\begin{align}
 \psi & \:\: = \; 
 k \; \ln\left[
\;\;\sum_{n=0}^{\infty}
\;\;\sum_{n'=0}^{\infty}
\; ( 2\:n' \:+\: 1 ) \:.\: 
\exp\left(-\:\frac{\overline{u}_{\,n\:n'}}{k\:T}\right)
\:\;\right]
\label{label_eq_19} \; .
\end{align}

% Die Bestimmung von $g$ und $g'$ habe ich in einer demnächst in den Annalen der Physik(1) erscheinenden Arbeit über "die physikalische Struktur des Phasenraums" durchgeführt und erlaube mir die Resultate hier zu benutzen, was um so leichter geht, als die Bezeichnungen hier ganz dieselben sind wie dort. Es ist nämlich: $$(20)$$. 
I have carried out the determination of $g$ and $g'$ in a work about ``\,the physical structure of phase space\,'' that will soon appear in the Annals of 
%- - - - - - - - - - - - - - - - - -
Physics$\,$\footnote{$\:$ 
%Voraussichtlich im Band 50, 1916.
Probably in volume 50, 1916 {\it\color{blue}(Indeed: 1916. Die physikalische Struktur des Phasenraumes [\,The physical structure of phase space\,], Annalen der Physik Leipzig, Vol.50, p.385-418 / P. Marquet)}.}
%- - - - - - - - - - - - - - - - - -
and allow me to use the results here, which is all the easier, as the names here are exactly the same as there. 
Namely it is: 
%====================================
\begin{equation}
\left.
\begin{aligned}
%-------------------------------------------------------
 g  & \: = \; 2\:m\:R\:q\: 
 \left[\: \sin(\alpha) \:-\: \alpha \: \cos(\alpha) \:\right]
   \;  \\
%-------------------------------------------------------
  g'  & \: = \; 2\:\pi\:m\:R\:q\: \cos(\alpha)
   \; 
%-------------------------------------------------------
\end{aligned}
\;\;
\right\} 
\label{label_eq_20} \; .
\end{equation}
%====================================
%\vspace*{0mm}
% Da die Energie $u=...$ sich nieht bequem durelr $g$ und $g'$ ausdrücken läßt, so führt die Berechnung von (19) für den allgemeinen Fall auf sehr komplizierte Ausdrücke, sie wird aber leicht, wenn man sich wieder auf die Betrachtung hoher oder tiefer Temperaturen beschränkt. 
Since the energy $u=m\:q^2/2$ cannot be conveniently expressed in terms of $g$ and $g'$, the calculation of (\ref{label_eq_19}) for the general case leads to very complicated expressions, 
%but it becomes easy if one considers again limited to considering high or low temperatures.
 but it becomes easy if one again restricts oneself to the consideration of high or low temperatures. 

% Bei hohen 'Temperaturen (großen Ordnungszahlen $n$ und $n'$) kann man $unn'$ ersetzen durch den Wert von u in einem beliebigen Punkte des Elementargebiets $(n\:n')$, und die Summation in (19) durch eine Integration. Dann ergibt sich mit Rücksicht auf (5), (6) und (7): $$AA$$.
% - - - - - - 
% At high temperatures (large ordinal numbers $n$ and $n'$), $unn'$ can be replaced by the value of u at any point of the elementary region $(n\:n')$, and the summation in (19) by an integration. Then, taking into account (5), (6) and (7), the result is: $$AA$$.
% - - - - - - 
\dashuline{At high temperatures} (large atomic numbers $n$ and $n'$) one can replace $\overline{u}_{\,n\:n'}$ by the value of $u$ at any point in the elementary domain $(n\:n')$, and {\it\color{blue}(replace)} the summation in (\ref{label_eq_19}) by an integration. 
Then, taking into account (\ref{label_eq_5}), (\ref{label_eq_6}) and (\ref{label_eq_7}), we get: 
\begin{align}
 \psi & \:\: = \; 
 k \; \ln\left[\;
\bigintsss_{\,0}^{\infty}\!\!\!
\bigintsss_{\,0}^{\infty}
\exp\left(-\:\frac{u}{k\:T}\right)
\:.\:\frac{d\,g \:.\: d\,g'^{\,2}}{h^3}
\:\;\right]
 \nonumber \; .  % \label{label_eq_6} \; .
\end{align}
% Zur Ausführung dieser Integration benutzen wir als Integrationsvariable $q$ und $\alpha$ statt $g$ und $g'**2$, vermittels der Gleichungen (20) und der daraus folgenden Beziehung: $$AA$$.
To carry out this integration, we can use $q$ and $\alpha$ as integration variables instead of $g$ and $g'^{\,2}$, by means of equations (\ref{label_eq_20}) and the following relationship: 
\begin{align}
 d\,g \:.\: d\,g'^{\,2}  
 & \: = \; 16\:\pi^2\:m^3\:R^3\:q^2\: 
  \sin^2(\alpha) \: \cos(\alpha)
  \; d\alpha \; dq 
 \nonumber \; .  % \label{label_eq_6} \; .
\end{align}
% Dann ergibt sich: 
Then it results:
\vspace*{-1mm}
\begin{align}
 \psi & \:\: = \; 
 k \; \left\{\;
\frac{3}{2}\;
\ln\left(\:
\frac{2\:\pi\:m\:k\:T}{h^2}
\:\right)
\;+\;
\ln\left(\:\frac{4}{3}\:R^3\:\pi\:\right)
\:\;\right\} 
 \label{label_eq_21} \; ,
\end{align}
% und daraus nach (2) die Energie: 
and from this, according to (\ref{label_eq_2}), the energy {\color{blue}$u=T^2\:\partial\psi/\partial T$} is:
\vspace*{-1mm}
\begin{align}
\boxed{\:
 u \; = \; \frac{3}{2} \: k \: T
\:}
 \nonumber %  \label{label_eq_14}
 \; , 
\end{align}
% ferner nach (3) die Entropie: 
and furthermore, according to (\ref{label_eq_3}), the entropy {\it\color{blue}$s=\psi+u/T=\psi+(3/2)\:k
=\psi+k\;(3/2)\:\ln(e)$ with $e=\exp(1)$} is:
\vspace*{-3mm}
\begin{align}
\boxed{\:
 s \; = \; 
 k \; \left\{\;
\frac{3}{2}\;
\ln\left(\:
\frac{2\:\pi\:e\:m\:k\:T}{h^2}
\:\right)
\;+\;
\ln\left(\:\frac{4}{3}\:R^3\:\pi\:\right)
\:\;\right\} 
\:}
 \label{label_eq_22} \; .
\end{align}
% Ein Vergleich mit (15) zeigt, daß für hohe "Temperaturen die Entropie eines einzelnen Atoms in einer Hohlkugel die nämliche ist wie in einem rechtwinkligen Parellelepiped von gleichem Volumen, und es wird nicht allzu gewagt sein, hieraus weiter zu schließen, daß dieser Satz ganz allgemein für jede beliebige Form des Hohlraums gültig ist. 
% - - - - - - - - -
% A comparison with (\ref{label_eq_15}) shows that at high temperatures the entropy of a single atom in a hollow {\it\color{blue}(empty)} sphere is the same as in a rectangular parallelepiped of the same volume, and it will not be too daring to further conclude from this that this theorem is complete is generally valid for any shape of the cavity.
% - - - - - - - - -
A comparison with (\ref{label_eq_15}) shows that for high temperatures the entropy of a single atom in a hollow {\it\color{blue}(empty)} sphere is the same as in a rectangular parallelepiped of the same volume, and it will not be too hazardous to conclude from this that this theorem is generally valid for any shape of the cavity {\it\color{blue}(or hollow (empty) 
% - - - - - - - - - - - - - - - - - - - - - - - - - - - -
space)}.$\,${\color{blue}\footnote{$\:${\it\color{blue}Note that the only difference between (\ref{label_eq_15}) and (\ref{label_eq_22}) are the second logarithm terms of pure constants independent of the true variables (i.e. the mass $m$ and the absolute temperature $T$) / P. Marquet.}.}} 
% - - - - - - - - - - - - - - - - - - - - - - - - - - - -

% Bei tiefen Temperaturen reduziert sich in dem Ausdruck (19) von $\Psi$ die Doppelsumme auf ihr erstes Glied, also: $$Psi=$$,  wobei die Nullpunktsenergie nach (18): $$(23)$$.
\dashuline{At low temperatures}, in the expression (\ref{label_eq_19}) of $\psi$ the double sum reduces to its first term, i.e. to: 
\begin{align}
 \psi \;=\; -\:\frac{\overline{u}_{\,0\:0}}{T} 
 \nonumber % \label{label_eq_16} % \nonumber 
 \; ,
\end{align}
where, according to (\ref{label_eq_18}), {\bf\dashuline{the zero point energy is}}: 
\begin{align}
\boxed{\:
 \overline{u}_{\,0\:0} 
 \; = \;
 \frac{1}{h^3}
\bigintsss_{\,0}^{\,h}\!\!\!
\bigintsss_{\,0}^{\,h^2}
\frac{m}{2} \:.\: q^2 \:.\: d\,g \:.\: d\,g'^{\,2}
\:}
%\nonumber \; .  
 \label{label_eq_23} 
 \; .
\end{align}
% Die Ausführung der Integration ist umständlich und zeigt durch einen Vergleich mit (17), daß die Nullpunktsenergie wesentlich abhängig ist von der Form des Hohlraums.
Carrying out the integration is complicated but shows, by comparison with (\ref{label_eq_17}), that {\bf\dashuline{the zero point energy}} is essentially dependent on the shape {\it\color{blue}(dimensions/volume)} of 
% - - - - - - - - - - - - - - - - - - - - - - - - - - - -
the cavity.$\,${\color{blue}\footnote{$\:${\it\color{blue}Note that, similarly, the second logarithms in (\ref{label_eq_14}) or (\ref{label_eq_21}) for the characteristic function $\psi$, and in (\ref{label_eq_15}) or (\ref{label_eq_22}) for the entropy $s$, are exactly the volume of the cavity: ``\,$a\:b\:c$\,'' and ``\,$(4/3)\:\pi\:R^3$'' for a rectangular parallelepiped and a sphere of radius $R$, respectively / P. Marquet.}.}} 
% - - - - - - - - - - - - - - - - - - - - - - - - - - - -

% \bigintssss \bigintsss \bigintss \bigints \bigint 
% \dashuline{\,dashing}  -{\it\color{blue}(change)}
%- - - - - - - - - - - -
% $\,$\footnote{$\:$.}  (\ref{label_eq_thermal_1})
% $\,$\footnote{$\:${\it\color{blue}()}.}

\vspace*{-2mm}
\begin{center}
------------------------------------------------------------------------------------
\end{center}
\vspace*{-5mm}

%========================================
\section{\underline{Second part. A large number of atoms with all coherent degrees} \\ \underline{of freedom.} (p.659-665)}
%========================================
\label{Section-3}
\vspace*{0mm}
% Zweiter Teil. 
% Second part.
% Eine große Anzahl von Atomen mit lauter kohärenten Freiheitsgraden. 
% A large number of atoms with all coherent degrees of freedom.
% A large number of atoms with all coherent degrees of freedom. 

\noindent §\:4.
If we now move on to considering a large number $N$ of point-like atoms, in order to get to know the structure of the phase space, we first have to ask again about the shape of the path that a phase point follows in this $6N$-dimensional space 
% describes the movement of atoms, 
  during the motion of the atoms,
and this question cannot be answered until we make a premise about the forces with which the atoms act on one another during a collision. 

% We therefore want to 
  We shall therefore 
first introduce the most obvious hypothesis that these forces absolutely obey the laws of classical mechanics, and also the further well-known hypothesis that the phase point comes arbitrarily close to every point on its energy surface 
$U = const.$ in the course of its movement. 

Then the elementary regions of the phase space are bounded by a single type of surfaces $g$, namely the surfaces of constant energy: 
\begin{align}
 U & \: = \;
 U_0\:,\;U_1\:,\;U_2\:,\; .\:.\:.\: \;U_n\:,\; .\:.\:.\:
 \nonumber 
%\label{label_eq_23} 
 \; 
\end{align}
%so that there is 
 There is therefore 
only a single series of atomic numbers $n$, and the $3N$ degrees of freedom of the system are all coherent with each other.
%there is therefore only a single series of ordinal numbers $n$, and the $3N$ degrees of freedom of the system are all coherent with each other.

% In unseren Formeln ist dann f=3N, und nach (5): 
In our formulae, $f=3\:N$, and according to (\ref{label_eq_5}): 
\begin{align}
 dG & \:\: = \; 
\bigintsss\!\!\!
\bigintsss\!\!\!
\bigintsss_{\;\;U}^{\:U+dU}
\!\!\!\!\!\!\!\!
.\,.\,.\;\: d\phi_1 \; d\phi_2 \:\:
.\,.\,.\:\: d\psi_1 \; d\psi_2 \:\:
.\,.\,.
\;=\;
d\left({g^{3\,N}}\right) 
\label{label_eq_24} \; .
\end{align}
% Das Elementargebiet (n) besitzt nach (6) und (4) die Größe: $$(25)$$, wo $$(26)$$. 
According to (\ref{label_eq_6}) and (\ref{label_eq_4}), the elementary region (n) has the size: 
\begin{align}
 G_{n+1} \:-\: G_{n} & 
 \; = \; 
 g^{3\:N}_{\,n+1} \:-\: g^{3\:N}_{\,n}
 \; = \;
 \left[\:
  (n\,+\,1)^{3\:N} \:-\: (n)^{3\:N}
 \:\right] \; h^{3\:N}
\label{label_eq_25} 
\; ,
\end{align}
where 
\vspace*{-2mm}
\begin{align}
 G_{n} & \; = \; (n\:h)^{3\:N}
\label{label_eq_26} 
\; .
\end{align}
% Nun wollen wir $N$ so groß voraussetzen, daß in der Differenz (25) der Subtrahend gegen den Minuend verschwindet, was stets und nur dann zutrifft, wenn $$n<<N (27)$$, d.h. wenn nur solche Glieder der Summe in (1) für die Bildung der charakteristischen Funktion $\Psi$ in Betracht kommen, deren Ordnungszahl $n$ von kleinerer Größenordnung ist als $N$. Diese Voraussetzung ist um so leichter erfüllt, je tiefer die Temperatur ist. 
% - - - - - - - - - - - - - - - -
% Now let us assume that $N$ is so large that in the difference (25) the subtrahend disappears with respect to the minuend, which is always and only true if $$n<<N (27)$$, i.e. if only those members of the sum in (1) whose ordinal number $n$ is of a smaller order of magnitude than $N$ can be considered for the formation of the characteristic function $\Psi$. The lower the temperature, the easier it is to fulfil this requirement. 
% - - - - - - - - - - - - - - - -
Now we want to assume that $N$ is so large that in the difference (\ref{label_eq_25}) the subtrahend {\it\color{blue}$(n)^{3\:N}$} against the minuend {\it\color{blue}$(n\,+\,1)^{3\:N}$} disappears, which is always and only true if 
\vspace*{-2mm}
\begin{align}
 n \; \ll \; N  
\label{label_eq_27} 
\; ,
\end{align}
% i.e. if only such members of the Sum in (\ref{label_eq_1}) can be considered for the formation of the characteristic function $\Psi$, whose atomic number $n$ is of smaller magnitude than $N$. 
  i.e. if only those members of the sum in (\ref{label_eq_1}) whose ordinal number $n$ is of a smaller order of magnitude than $N$ can be considered for the formation of the characteristic function $\Psi$.
The lower the temperature, the easier it is to fulfill this requirement.

% Dann geht die Gleichung (25) über in: $$(28)$$.
Then equation (\ref{label_eq_25}) turns into: 
\vspace*{-2mm}
\begin{align}
 G_{n+1} \:-\: G_{n} & 
 \; = \;
  (n\,+\,1)^{3\:N} 
 \; h^{3\:N}
\label{label_eq_28} 
\; .
\end{align}
% Daraus folgt nach (7): 
From this it follows from (\ref{label_eq_7}):
\vspace*{-3mm}
\begin{align}
 p_{n} & 
 \; = \;
  (n\,+\,1)^{3\:N} 
\label{label_eq_29} 
\; 
\end{align}
% und nach (8) mit demselben Grade der Annäherung: 
and according to (\ref{label_eq_8}) with the same degree of approximation:
\begin{align}
\overline{U}_{n}
 & \:\: = \; 
\frac{\displaystyle
\bigintsss_{\,n}^{\,n+1}
U \:.\:
d\,G 
}{G_{n+1} \:-\: G_{n} } 
\;=\; {U}_{n+1}
\label{label_eq_30} \; ,
\end{align}
% und nach (8) mit demselben Grade der Annäherung: $$(30)$$, d. h. die mittlere Energie in dem von den Flächen $U= U_n$, und $U— U_{n+1}$ begrenzten Elementargebiet $(n)$ ist bis auf verschwindend kleines gleich der Energie $U_{n+1}$.
%  i.e. the average energy in the elementary region $(n)$ bounded by the areas $U= U_n$, and $U-U_{n+1}$ is equal to the energy $U_{n+1}$ up to vanishingly small.
% and according to (8) with the same degree of approximation: $$(30)$$, 
% i.e. the mean energy in the elementary region $(n)$ bounded by the surfaces $U= U_n$, and $U=U_{n+1}$ is equal to the energy $U_{n+1}$ except for a vanishingly small one.
  i.e. the average energy in the elementary region $(n)$ bounded by the areas $U=U_n$ and $U=U_{n+1}$ is equal to the energy $U_{n+1}$, up to vanishingly small.

% Dabei ist $U$, nach (24) und (26) bestimmt durch die Beziehung: (31).
Here $U$, according to (\ref{label_eq_24}) and (\ref{label_eq_26}), is determined by the relation: 
\begin{align}
 G_n & \:\: = \; 
\bigintsss\!\!\!
\bigintsss\!\!\!
\bigintsss_{\;\;U=0}^{\:U=U_n}
\!\!\!\!\!\!\!\!
.\,.\,.\;\: d\phi_1 \; d\phi_2 \:\:
.\,.\,.\:\: d\psi_1 \; d\psi_2 \:\:
.\,.\,.
\;=\;
(n\:h)^{3\:N}
\label{label_eq_31} \; .
\end{align}
% Diese Werte, in (1) eingesetzt, ergeben die charakteristische Funktion des Körpers: 
These values, inserted in (\ref{label_eq_1}), result in the characteristic function of the body:
\begin{align}
 \Psi &
 \: = \; 
 k \; \ln\left[
\;\;\sum_{n=0}^{\infty}
\; (n\,+\,1)^{3\:N}  \:.\:\: 
\exp\left(-\:\frac{U_{\,n+1}}{k\:T}\right)
\:\;\right]
 \; \approx \; 
 k \; \ln\left[
\;\;\sum_{n=0}^{\infty}
\; (n\,)^{3\:N}  \:.\:\: 
\exp\left(-\:\frac{U_{\,n}}{k\:T}\right)
\:\;\right]
\label{label_eq_32} \; .
\end{align}
% Wie man sieht, beziehen sich jetzt die Ordnungszahlen nicht mehr auf die Elementargebiete, sondern auf die Grenztlächen der Elementargebiete, und damit rechtfertigt sich die am Schluß des §1 ausgesprochene Behauptung, daß es hier keinen wesentlichen Unterschied macht, ob man die Phasenpunkte im Innern der Elementargebiete befindlich oder an ihren Grenzen angehäuft annimmt. Dies wird überhaupt immer dann der Fall sein, wenn es sich um sehr viele kohärente Freiheitsgrade handelt.
%- - - - - - - - - - - -
% Deepl: 
As can be seen, the ordinal numbers now no longer refer to the elementary domains, but to the boundary surfaces of the elementary domains, and this justifies the assertion made at the end of §\,1 that it makes no essential difference here whether one assumes the phase points to be located inside the elementary domains or accumulated at their boundaries. This will always be the case when there are many coherent degrees of freedom.
%- - - - - - - - - - - -
%Google: As you can see, the ordinal numbers no longer refer to the elementary areas, but to the boundary areas of the elementary areas, and this justifies the assertion made at the end of §1 that it makes no essential difference here whether the phase points are inside the Elementary areas located or clustered at their borders. This will always be the case when there are very many coherent degrees of freedom.

% \bigintssss \bigintsss \bigintss \bigints \bigint 
% \dashuline{\,dashing}  -{\it\color{blue}(change)}
%- - - - - - - - - - - -
% $\,$\footnote{$\:$.}  (\ref{label_eq_31})
% $\,$\footnote{$\:${\it\color{blue}()}.}

% \bigintssss \bigintsss \bigintss \bigints \bigint 
% \dashuline{\,dashing}  -{\it\color{blue}(change)}
%- - - - - - - - - - - -
% $\,$\footnote{$\:$.}  (\ref{label_eq_thermal_1})
% $\,$\footnote{$\:${\it\color{blue}()}.}

\vspace*{-4mm}
\begin{center}
---------------------------------------------------
\end{center}
\vspace*{-3mm}

\noindent §\:5. 
In the above calculations, we have made use of a prerequisite which has not been explicitly emphasised anywhere so far, but which is nevertheless essential for the validity of the derived formulae: namely that all $N$ atoms of the body under consideration are dissimilar. 
Indeed, only in this case does a certain point in phase space correspond to each physical state of the body defined in a microscopically precise sense. If, however, the body contains groups of atoms of the same type, this is no longer the case. 
Instead, a more or less large number of physically completely equivalent points in the phase space are assigned to a certain physical state of the body, since a certain point in the phase space requires certain coordinates and velocities for each individual atom. 
As many permutations as the atoms of the same type allow, just as many phase points correspond to a certain physical state. 
For ease of expression, I will therefore distinguish between ``\,phase point\,'' and ``\,state point.\,'' 

% Sind von den N Atomen $N...$ unter sich gleichartig, so entsprechen jedem Zustandspunkt $N... = R$ Phasenpunkte, und sowohl der ganze Phasenraum wie auch jedes Elementargebiet des Phasenraumes zerfällt in $R$ physikalisch vollkommen kongruente Stücke, von denen man ein beliebiges herausgreifen und als Repräsentant des »Zustandsraumes« bzw. eines »Zustandsgebiets« betrachten kann. Die Größe eines Zustandsgebiets ist der $R$te Teil des entsprechenden Phasengebiets. 
%- - - - - - - - - - - -
% Deepl: If $N...$ of the N atoms are identical among themselves, then each state point $N... = R$ phase points, 
% and the entire phase space as well as each elementary region of the phase space decomposes into $R$ physically completely congruent pieces, 
% of which one can pick out any one and regard it as a representative of the "state space" or a "state region". 
%The size of a state region is the $R$th part of the corresponding phase region. 
%- - - - - - - - - - - -
%Google: If the N atoms $N...$ are similar, then each state point $N... = R$ corresponds to phase points, 
% and both the entire phase space and each elementary region of the phase space break down into $R$ physically completely congruent pieces, 
% Any one of which can be picked out and viewed as a representative of the “state space” or a “state area”. 
%The size of a state domain is the $R$th part of the corresponding phase domain.
%- - - - - - - - - - - -
If $N_1,\:N_2,\:N_3, \,.\,.\,.$ of the $N$ atoms are identical, then each state point 
$N_1!\:\:N_2!\:\:N_3! \,.\,.\,. = {\cal R}$ 
corresponds to phase points, 
and the entire phase space (as well as each elementary region of the phase space) decomposes into ${\cal R}$ physically completely congruent pieces,
of which one can pick out any one and regard it as a representative of the ``\,state space\,'' or a ``\,state region.\,'' 
The size of a state region is the ${\cal R}\,$th part of the corresponding phase region. 

The question now arises as to whether and which modifications are to be made in this case to the above equations to determine the thermodynamic characteristic function. 
This question cannot be decided a priori (just as the thermodynamic probability $W$ cannot be derived a priori), but can only be answered by making a determination that is as generally useful and as plausible as possible, trusting in the feasibility of an appropriate theory, which leads to correct results in all controllable cases.
The following sentence should be understood in this sense, which, as far as I can see so far, generally solves the problem at hand.

% Wenn unter den $N$ Atomen des Körpers $N....$ unter sich gleichartig sind, so bleiben die Gleichungen (29), (30) und (32) ganz ungeändert. In der Tat behalten ja die Größen $pn$ und $Un$ ihre Bedeutung, ob man sie nun auf den Phasenraum oder auf den Zustandsraum bezieht, da beide Größen Verhältniszahlen vorstellen, deren Wert von einer gleichmäßigen Veränderung des Zählers und Nenners nicht beeinflußt werden kann. Dagegen ändert sich die zur Bestimmung der Größe von $Un$ dienende Gleichung (31) insofern, als die ihr zugrunde liegende Beziehung (26) nicht für den Phasenraum, sondern für den Zustandsraum Gültigkeit besitzt. Wenn wir also nach wie vor die Gebietsgröße $Gn$ auf den Phasenraum beziehen, so ist dieselbe im Falle gleichartiger Atome $\R$mal so groß anzunehmen wie der in (26) angegebene Wert, und die Gleichung (31) geht über in: $$(33)$$.
If among the $N$ atoms of the body $N_1,\:N_2,\:N_3, \,.\,.\,.$  are similar, then equations (\ref{label_eq_29}), (\ref{label_eq_30}) and (\ref{label_eq_32}) remain completely unchanged. 
In fact, the quantities $p_n$ and $\overline{U}_n$ retain their meaning whether one refers to the phase space or to the state space, since both quantities represent ratios whose value cannot be influenced by a uniform change in the numerator and denominator. 
On the other hand, equation (\ref{label_eq_31}), which is used to determine the size of $U_n$, changes in that the relationship (\ref{label_eq_26}) on which it is based is not valid for the phase space, but for the state space. 
If we continue to relate the area size $G_n$ to the phase space, then in the case of similar atoms it can be assumed to be ${\cal R}$ times as large as the value given in (\ref{label_eq_26}), and equation (\ref{label_eq_31}) turns into: 
\vspace*{-1mm}
\begin{align}
 G_{n} & \; = \; 
         N_1!\:\:N_2!\:\:N_3! \; \,.\,.\,. \; 
         (n\:h)^{3\:N}
\label{label_eq_33} 
\; .
\end{align}
% Für lauter gleichartige Atome erhält man spezieller: 
More specifically, for all similar atoms one obtains:
\begin{align}
 G_n & \:\: = \; 
\bigintsss\!\!\!
\bigintsss\!\!\!
\bigintsss\!\!\!
\bigintsss_{\;\;U=0}^{\:U=U_n}
\!\!\!\!\!\!\!\!
.\,.\,.\;\: d\phi_1 \; d\phi_2 \:\:
.\,.\,.\:\: d\psi_1 \; d\psi_2 \:\:
.\,.\,.
\;=\; N\,! \;(n\:h)^{3\:N}
\label{label_eq_34} \; .
\end{align}

% ``\, \,''  (\ref{label_eq_24})  
% \bigintssss \bigintsss \bigintss \bigints \bigint 
% \dashuline{\,dashing}  -{\it\color{blue}(change)}
%- - - - - - - - - - - -
% $\,$\footnote{$\:$.}  (\ref{label_eq_thermal_1})
% $\,$\footnote{$\:${\it\color{blue}()}.}

\vspace*{-4mm}
\begin{center}
---------------------------------------------------
\end{center}
\vspace*{-3mm}

\noindent §\:6. 
% Die weitere Rechnung beschränken wir auf den Fall eines aus $N$ gleichartigen Atomen mit der Masse $m$ bestehenden Gases. Dann ist die Energie, bei gleicher Bezeichnung wie in §\,2: $$(35)$$,  und die Grenzwerte $U_n$ der Energie bestimmen sich nach (34) durch die Beziehung: $$(35a)$$.
We restrict the further calculation to the case of a gas consisting of $N$ similar atoms with mass $m$. 
Then the energy, with the same designation as in §\,2: 
\begin{align}
 U & \: = \; 
 \frac{1}{2\:m}\:\:
 \sum_{k=1}^{N}\:
 \left(\:
   \xi_k^2 \:+\: \eta_k^2 \:+\:\zeta_k^2 
 \:\right)
\label{label_eq_35} \; ,
\end{align}
and the limit values $U_n$ of the energy are determined according to (\ref{label_eq_34}) by the relationship: 
\begin{align}\tag{35a}
 G_n & \:\: = \; 
\bigintsss\!\!\!
\bigintsss\!\!\!
\bigintsss_{\;\;U=0}^{\:U=U_n}
\!\!\!\!\!\!\!\!
.\,.\,.\;\: dx_k \; dy_k \; dz_k \:\:
.\,.\,.\;\: d\xi_k \; d\eta_k \; d\zeta_k \:\:
.\,.\,.
\;=\; N\,! \;(n\:h)^{3\:N}
\label{label_eq_35a} \; .
\end{align}

% Die Integration ist zu erstrecken für die Raumkoordinaten jedes Atoms über das ganze Volumen $V$ des Gases, für die Impulskoordinaten über alle Kombinationen, welche einer Gesamtenergie des Gases zwischen  $0$ und $U_n$ entsprechen. Dieses Integrationsgebiet läßt sich nach (35) auffassen als das Volumen einer Kugel in einem $3\:N$-dimensionalen Raume, vom Radius $V2mUT$, welches gleich ist  $$AA$$.
%- - - - - - - - - - - -
The integration must be extended for the spatial coordinates of each atom over the entire volume $V$ of the gas, and for the momentum coordinates over all combinations that correspond to a total energy of the gas between $0$ and $U_n$. According to (\ref{label_eq_35}), this integration domain can be understood as the volume of a sphere in a $3\,N$-dimensional space, of radius $\sqrt{\,2\:m\:U_n\,}$, which is equal to 
\begin{align}
 \frac{2\:\pi^{\,3\,N/2}}
      {\displaystyle 3\:N\:.\:
      \left(\frac{3\:N}{2}\:-\:1 \right)!}
 \:\:.\:\:
 (\,2\:m\:U_n\,)^{3\,N/2}
%\label{label_eq_35a} 
\nonumber
\;\: .
\end{align}
% Daraus folgt: $$A$$,  und, mit Benutzung des STIRLINGschen Satzes, bei Fortlassung kleinerer Glieder: $$(36)$$.
From this it follows: 
\vspace*{-3mm}
\begin{align}
 V^{N} \:.\:\:
 \frac{2\:\pi^{\,3\,N/2}}
      {\displaystyle 3\:N\:.\:
      \left(\frac{3\:N}{2}\:-\:1 \right)!}
 \:\:.\:\:
 (\,2\:m\:U_n\,)^{3\,N/2}
 \;=\; N\,! \;(n\:h)^{3\:N}
%\label{label_eq_35a} 
\nonumber
\;\: ,
\end{align}
and, using Stirling's theorem 
{\it\color{blue}$\ln(N!) \approx N\:\ln(N) - N$}, 
if smaller terms are omitted: 
\begin{align}
 U_n \;=\;
 \frac{3\:n^2\:h^2\:N^{5/3}}
      {4\:\pi\:e^{5/3}\:m\;V^{2/3}}
\label{label_eq_36} 
%\nonumber
\;\: .
\end{align}

% Bezeichnet man zur Abkürzung die durch die Atomzahl $N$ dividierten Werte des Volumens $V$, der Energie $U$, der charakteristischen Funktion $\Psi$, der Entropie $S$, mit den entsprechenden kleinen Buchstaben $v,u,\psi,s$, und setzt ferner die Zahl: $$(37)$$,  so folgt aus (32) als charakteristische Funktion: $$(38)$$,  ferner aus (2) als Atomenergie:  $$(39)$$  und aus (3) als Atomentropie:  $$(40)$$.
%- - - - - - - - - - - -
For abbreviation, let us denote with the corresponding small letters $v, u,\psi, s$ the values of the volume $V$, the energy $U$, the characteristic function $\Psi$ and the entropy $S$ divided by the atomic number $N$, and further sets the number: 
\vspace*{-3mm}
\begin{align}
  \tau \;=\;
 \frac{h^2}
      {4\:\pi\:e^{5/3}\:m\;v^{2/3}\:k\:T}
\label{label_eq_37} 
%\nonumber
\;\: ,
\end{align}
then from (\ref{label_eq_32}) it follows as a characteristic function: 
\begin{align}
 \psi &
 \: = \; 
 \frac{k}{N} \:.\:
\ln\left[
\;\;\sum_{n=0}^{\infty}
\; \left[\:
       n\:.\:\exp\left(\:-\,\tau\:n^2\:\right)
   \:\right]^{3\:N}
\:\;\right]
\label{label_eq_38} \; ,
\end{align}
further from (\ref{label_eq_2}) {\color{blue} and
from (\ref{label_eq_37}) leading to 
$u=T^2\:\partial\psi/\partial T 
  = -\,(T\:\tau)\:\partial\psi/\partial \tau$,}
as an atomic energy: 
\vspace*{0mm}
\begin{align}
 u &
 \: = \; 
 3\;k\;T\;\tau 
 \:\:.\:\;
 \frac{\displaystyle
    \sum_{n=0}^{\infty}
    \;n^2 \:.
    \;\left[\:
       n\:\:\exp\left(\:-\,\tau\:n^2\:\right)
    \:\right]^{3\:N}}
      {\displaystyle
    \sum_{n=0}^{\infty}
    \;\left[\:
       n\:\:\exp\left(\:-\,\tau\:n^2\:\right)
    \:\right]^{3\:N}}
\label{label_eq_39} 
\; ,
\end{align}
and from (\ref{label_eq_3}) as an atomic entropy: 
\vspace*{-3mm}
\begin{align}
 s & \: = \; \psi \;+\; \frac{u}{T}
\label{label_eq_40} 
\; .
\end{align}
% Für hohe Temperaturen ($\tau \ll 1$) kann man $n$ durchweg als groß betrachten und die Summe in (38) als (Eulersches) Integral schreiben; also: 
For \dashuline{high temperatures} 
($\tau \;{\color{blue}\propto 1/T\,} \ll 1$) 
one can always consider $n$ to be large and write the sum in (\ref{label_eq_38}) as an (Eulerian) integral, and therefore:
\vspace*{-1mm}
\begin{align}
 \psi & \: = \; \frac{k}{N} \:.\:
  \ln\left[
    \;\;
    \bigintsss_{\;\;0}^{\:\infty}
    \;\left[\:
       n\:\:\exp\left(\:-\,\tau\:n^2\:\right)
    \:\right]^{3\:N}
    \:.\: dn
\:\;\right]
\label{label_eq_41} 
\; , \\
 \psi & \: = \; -\:\frac{3}{2} \:\: k
      \; \ln\left( \, 2 \; e \; \tau \, \right)  
 {\color{blue} 
 \:\; = \; \frac{3}{2} \:\: k
   \; \ln\left( \, 
   \frac{2\:\pi\:m\:k\:T \:(e\:v)^{2/3}}{h^2}
   \, \right)      
 }
\label{label_eq_42} 
\; .
\end{align}
% Dieses Resultat läßt sich auch auf viel einfacherem Wege erhalten. wenn man bedenkt, daß wegen der Größe von N das größte Glied der Summe in (38), nämlich das Glied mit der Ordnungszahl $n'=???$ (wegen der Kleinheit von $\tau$ als ganze Zahl anzusehen) alle anderen Glieder derart an Größe überragt, daß der Wert der ganzen unendlichen Summe sich auf dieses einzige Glied reduziert. Dann folgt aus (38) für $\psi$ direkt der Wert (42), ferner aus (2) oder aus (39): $$(43)$$ und nach (40) und (37): $$(44)$$.
%- - - - - - - - - - - -
% Google: 
% ​This result can also be obtained in a much simpler way. if one considers that because of the size of N, the largest term of the sum in (38), namely the term with the ordinal number $n'=???$ (to be viewed as an integer because of the smallness of $\tau$), all others The members are so great in size that the value of the entire infinite sum is reduced to this single term. Then from (38) the value (42) follows directly for $\psi$, further from (2) or from (39): $$(43)$$ and from (40) and (37): $$(44)$$.
%- - - - - - - - - - - -
% DeepL:
This result can also be obtained in a much simpler way if one considers that, because of the size of $N$, the largest member of the sum in (\ref{label_eq_38}) --namely the member with the ordinal number $n'=1/\sqrt{2\:\tau}$ {\it\color{blue}computed with 
$\partial/\partial\,n'\,[\:n'\,\exp(-\tau\,n'^{\,2})\:]=0\,$}, 
to be regarded as an integer because of the smallness of $\tau$-- exceeds all other members in size in such a way that the value of the whole infinite sum is reduced to this single member. 
% Then the value (42) follows directly from (38) for $\psi$, furthermore from (2) or from (39): $$(43)$$ and after (40) and (37): $$(44)$$.
Therefore, {\it\color{blue}the characteristic function} $\psi$ given by (\ref{label_eq_42}) follows directly from 
%- - - - - - - - - - - -
(\ref{label_eq_38}){\it\color{blue}$\,$\footnote{$\:${\it\color{blue}Indeed, with the sole term $n=n'=1/\sqrt{2\:\tau}$ in the sum  (\ref{label_eq_38}), we get 
$\psi = (k/N)\:.\:
\ln\left[\:
n'\,\exp(-\tau\,n'^{\,2})
\:\right]^{3 N}
= (k/N)\:.\:(3\:N)\:
\left[\:\ln(n')\,-\,\tau\:n'^{\,2}\:\right]$, 
which leads to (\ref{label_eq_42}) with 
$\ln(n')=(-1/2)\:\ln(2\,\tau)$ and
$-\,\tau\:n'^{\,2} = -\,(1/2)\,\ln(e)$ / P. Marquet}.}}.
%- - - - - - - - - - - -

% ``\, \,''  (\ref{label_eq_24})  \vspace*{-3mm}
% \bigintssss \bigintsss \bigintss \bigints \bigint 
% \dashuline{\,dashing}  -{\it\color{blue}(change)}
%- - - - - - - - - - - -
% $\,$\footnote{$\:$.}  (\ref{label_eq_thermal_1})
% $\,$\footnote{$\:${\it\color{blue}()}.}

Furthermore {\it\color{blue}the internal energy follows} from (\ref{label_eq_2}) {\color{blue} and $u=T^2\:\partial\psi/\partial T$ applied to (\ref{label_eq_42}) with $\tau \propto 1/T $ given by (\ref{label_eq_37}),} or {\it\color{blue}directly\,} from (\ref{label_eq_39}){\it\color{blue}, to give\,}:  
\vspace*{-1mm}
\begin{align}
\boxed{\:
 u \; = \; \frac{3}{2} \: k \: T
\:}
 \label{label_eq_43}
 \; , 
\end{align}
and {\it\color{blue}the entropy follows from} (\ref{label_eq_40}) and (\ref{label_eq_37}){\it\color{blue}, leading to\,}: 
\vspace*{-1mm}
\begin{align}
 s & \: = \; 
 {\it\color{blue}\psi \:+\: \frac{u}{T} \;=\;\:}
 -\:\frac{3}{2} \:\: k
      \; \ln\left( \, 2 \; \tau \, \right)  
   \; , \;\quad \mbox{or:} \;\;\;
\boxed{\:
 s \; = \; 
\frac{3}{2}\; k \;
\ln\left(\:
\frac{2\:\pi\:e^{5/3}\:m\:v^{2/3}\:k\:T}{h^2}
\:\right)
\:} 
\label{label_eq_44} 
\; .
\end{align}
%- - - - - - - - - - - -

{\it \color{blue} \noindent 
P. Marquet: Note that the relationship (\ref{label_eq_44}) may be put into a formulation similar to (\ref{label_eq_15}) or (\ref{label_eq_22}), with as expected the second logarithm term only depending on a pure geometric value --\,made of the product of the pure constant term $e=\exp(1)$ by the volume $(v)$, leading to:
%-----------------------------------------------------------
\vspace*{0mm}
\begin{align}\tag{44bis}
{\:
 s\,(T,v,m)  \; = \; 
 k \; \left\{\;
\frac{3}{2}\;
\ln\left(\:
\frac{2\:\pi\:e\:m\:k\:T}{h^2}
\:\right)
\;+\;
\ln\left(\: e \: v \:\right)
\:\;\right\} 
\:}
\nonumber\; .
\label{label_eq_44bis} 
\end{align}
%-----------------------------------------------------------
Moreover, it is possible to rewrite (\ref{label_eq_44}) to arrive at what is nowadays called the ``\,Sackur-Tetrode\,'' formulation for the absolute entropy of a monoatomic gas, namely any of the equivalent formulas: 
%-----------------------------------------------------------
\vspace*{0mm}
\begin{align}\tag{44ter}
{\:
 S\,(T,v,m) \; = \;
 N \:.\; s \;=\; 
 R \;\,.\: 
\ln\left[\:
\frac{\:e^{5/2}\;\,v\:\left(\,2\:\pi\:m\:k\:T\,\right)^{3/2}\,}
     {h^3}
\:\right]
\:}
\nonumber\; ,
\label{label_eq_44ter} 
\end{align}
%-----------------------------------------------------------
\vspace*{0mm}
\begin{align}\tag{44ter-T-V}
{\:
 S\,(T,V,M) \; = \;
 R \;\,.\: 
\ln\left\{\:
 \frac{5}{2}
 \:+\:
 \frac{3}{2} \:
 \ln(T)
 \:+\:
 \ln(V)
 \:+\:
 \frac{3}{2} \:
 \ln(M)
 \:+\:
 \ln\left[\:
 \frac{\:\left(\,2\:\pi\:k\,\right)^{3/2}\,}
      {h^3\;N^{5/2}}
 \:\right]
\:\right\}
\:}
\nonumber\; ,
\label{label_eq_44ter_v} 
\end{align}
%-----------------------------------------------------------
\vspace*{0mm}
\begin{align}\tag{44ter-T-P}
 S\,(T,P,M) \; = \;
 R \;\,.\: 
\ln\left\{\:
 \frac{5}{2}
 \:+\:
 \frac{5}{2} \:
 \ln(T)
 \:-\:
 \ln(P)
 \:+\:
 \frac{3}{2} \:
 \ln(M)
 \:+\:
 \ln\left[\:
 \frac{\:\left(\,2\:\pi\,\right)^{3/2}\;k^{5/2}\,}
      {h^3\;N^{3/2}}
 \:\right]
\:\right\}
\nonumber\; ,
\label{label_eq_44ter_P} 
\end{align}
%-----------------------------------------------------------
where $N$ is the (atomic) Avogadro's number, $R = N \: k \approx 8.314$ the gas constant, $M=N\:m$ the molar mass, $V=N\:v$ the molar volume and $P\:V=R\:T$ the molar gas equation.
the last logarithm term in (\ref{label_eq_44ter_P}) has the value of about $18.222\,85$~SI-unit.
}
\noindent
This expression for the entropy of an ideal monatomic gas {\it\color{blue}given by (\ref{label_eq_44}) or (\ref{label_eq_44bis})\,}, which differs from the value derived above in (\ref{label_eq_22}) for the entropy of a single atom only by {\bf\dashuline{an additive member}}, is completely identical with the expression determined by 
%- - - - - - - - - - - -
\dashuline{Stern}$\,$\footnote{$\:$O. Stern, Phys. Zeitschr. 14, S. 629, 1913.} 
%- - - - - - - - - - - -
and von \dashuline{Tetrode}$\,$\footnote{$\:$H. Tetrode, Ann. d. Phys. 38, S. 434, 1912. Ber. d. Akad. d. Wiss. v. Amsterdam, 27. Februar und 27. März, 1915, Gleichung{\it\color{blue}(equation)} (16).} 
%- - - - - - - - - - - -
in quite different ways and well confirmed by experience.
Differently, %  while 
%- - - - - - - - - - - -
\dashuline{Sackur}$\,$\footnote{$\:$O. Sackur, Ann. d. Phys. 36, S. 958, 1911 ; 40, S. 67, 1913.}, 
%- - - - - - - - - - - -
who was the first to calculate the absolute entropy of a gas from the assumption of finite elementary regions of probability, obtained the value (\ref{label_eq_22}), 
while {\bf\dashuline{the entropy constant}} in the expressions of 
%- - - - - - - - - - - -
\dashuline{Sommerfeld}$\,$\footnote{$\:$A. Sommerfeld, Göttinger Vorträge über die kinetische Theorie der Materie und der Elektrizität {\it\color{blue}(Göttingen lectures on the kinetic theory of matter and electricity)} 1914, S. 125 (B. G. Teubner).} 
%- - - - - - - - - - - -
and \dashuline{Keesom}$\,$\footnote{$\:$H. Keesom, Phys. Zeitschr. 14, S. 665, 1913; 15, S. 695, 1914.} 
%- - - - - - - - - - - -
show even greater deviations. 

% Dagegen ist neuerdings Nernst durch seine Hypothese der Nullpunktsstrahlung zu einem von dem Tetrodeschen Wert nur wenig verschiedenen Wert der Entropiekonstanten geführt worden, und zwar gilt die Nernstsche Berechnung, gegenüber denjenigen von STERN und von TETRODE, nicht nur für den idealen Gaszustand, sondern auch für beliebig tiefe Werte der Temperatur. Es wird nämlich hiernach die gesamte Energie des Gases, einschließlich der Nullpunktsenergie, dargestellt durch den Ausdruck: $$(45)$$,  wobei zur Abkürzung gesetzt ist: $$(46)$$. 
%- - - - - - - - - - - -
% Google:
% On the other hand, Nernst's hypothesis of zero-point radiation has recently led him to a value of the entropy constant that is only slightly different from Tetrode's value. Nernst's calculation, compared to those of STERN and TETRODE, applies not only to the ideal gas state, but also to any arbitrary state low temperature values. The total energy of the gas, including the zero point energy, is represented by the expression: $$(45)$$, where the abbreviation is: $$(46)$$.
%- - - - - - - - - - - -
% Deepl:
In contrast, 
%- - - - - - - - - - - -
\dashuline{Nernst}'s$\,$\footnote{$\:$W. Nernst. Verhandl. d. Deutschen Phys. Ges. 18, S. 83. 1916.}
%- - - - - - - - - - - -
{\bf\dashuline{hypothesis of zero-point radiation}} has recently led to a value of {\bf\dashuline{the entropy constant}} that differs only slightly from Tetrode's value, and Nernst's calculation, in contrast to those of Stern and Tetrode, is valid not only for the ideal gas state, but also for arbitrarily low values of temperature. 
According to this {\it\color{blue}\dashuline{Nernst}'s hypothesis}, the total energy of the gas, including {\bf\dashuline{the zero-point energy}}, is represented by the expression:
\vspace*{0mm}
\begin{align}
 \mbox{{\it\color{blue}(Nernst's value:)}}
 \quad\quad
 \frac{U}{N} \;=\; 
 u \; = \;
 \frac{3}{2} \: 
 \frac{h\:\nu}
 {\displaystyle
 \exp\left( \frac{h\:\nu}{k\:T} \right) \:-\: 1}
 \;+\; 
 \frac{3}{2} \: h \: \nu
 \label{label_eq_45}
 \; , 
\end{align}
whereby the abbreviation {\it\color{blue}(for the frequency $\nu$)} is: 
\vspace*{-1mm}
\begin{align}
 \mbox{{\it\color{blue}(Nernst's value:)}}
 \quad\quad
 \nu \; = \;
 \frac{h}
 {4\;\pi\;m\;v^{2/3}}
 \label{label_eq_46}
 \; . 
\end{align}

% Der Vergleich mit dem von mir gefundenen Ausdruck (39) ergibt folgendes: Für hohe Temperaturen stimmen die Werte überein, wie natürlich. Für tiefere Temperaturen ergibt sich nach (39) die Nullpunktsenergie als: $$(47)$$, dagegen nach (45):  $$(48)$$, also im Verhältnis $e... = 5.3:2$ größer. 
%- - - - - - - - - - - -
% DeepL: The comparison with the expression I found (39) shows the following: For high temperatures the values agree, as is natural. For lower temperatures, (39) gives the zero-point energy as: $$(47)$$, whereas (45) gives it as: $$(48)$$, i.e. in the ratio $e... = 5.3:2$ greater. 
%- - - - - - - - - - - -
% Google:
% The comparison with the expression (39) I found results in the following:
The comparison {\it\color{blue}of (\ref{label_eq_45})} with the expression (\ref{label_eq_39}) shows the following: \\
- for \dashuline{high temperatures}, the values agree, as is natural; and  \\
- for \dashuline{lower temperatures}, according to (\ref{label_eq_39}), 
  {\bf\dashuline{the zero-point energy}} 
%- - - - - - - - - - - -
  is:{\color{blue}$\,$\footnote{$\:${\it\color{blue}Note that Max Planck made a typo in the numerator of (\ref{label_eq_47}), with ``\,$h$\,'' instead of ``\,$h^2$\,'' / P. Marquet}.}} 
%- - - - - - - - - - - -
\vspace*{0mm}
\begin{align}
 \boxed{\;
 u_0 
 \; = \; 3 \; k \; T \: \tau
 \; = \; 
 \frac{3 \: h^{{\color{blue}2}}}
 {4\;\pi\;e^{5/3}\;m\;v^{2/3}}
 \;}
 \label{label_eq_47}
 \; , 
\end{align}
\hspace*{2mm} whereas, according to {\it\color{blue}\dashuline{the Nernst}'s hypothesis leading to} (\ref{label_eq_45}) 
{\it\color{blue}and for $T \approx 0$, the first term  \\
\hspace*{2mm}
vanishes and thus $u_0 \approx (3/2)\:h\:\nu$,
with $\nu$ given by (\ref{label_eq_46}) leading to}: 
\vspace*{0mm}
\begin{align}
 \mbox{{\it\color{blue}(Nernst's value:)}}
 \quad\quad
 u_0 
 \; = \; 
 \frac{3 \: h^2}
 {8\;\pi\;m\;v^{2/3}}
 {\color{blue}
 \:\;=\;
 \frac{3 \: h^{{\color{blue}2}}}
 {4\;\pi\;e^{5/3}\;m\;v^{2/3}}
 \; \times \;
 \frac{e^{5/3}}{2}
 }
 \label{label_eq_48}
 \; , \quad\quad\quad\quad
\end{align}
\hspace*{2mm}
which is larger {\it\color{blue}than (\ref{label_eq_47})} in the ratio 
%- - - - - - - - - - - -
$e^{5/3}/2 \approx 2.65$. % $\,$\footnote{$\:${\it\color{blue}However, we know today that the relationship (\ref{label_eq_45}) suggested by Nernst for the atomic internal energy $u=U/N$ was not valid / P. Marquet}.}.
%- - - - - - - - - - - -

{\it \color{blue} \noindent 
P. Marquet: the proof of (\ref{label_eq_47}) is obtained by computing the sums over 
$n=0$,\:$1$, \:$2$, \:$3$, \:.\:.\:.\:, 
\:$n$, \:.\:.\:.\: (up to \:$+\infty$\,) 
both at the numerator and the denominator of
(\ref{label_eq_39}), to give: 
%====================================
\begin{equation}\tag{39bis}
\left.
\begin{aligned}
%-------------------------------------------------------
 u & \: = \; (\,3\:k\:T\:\tau\,) 
 \:.\:
 \frac{\displaystyle
    0  
  \!+\left(e^{-\tau}\right)^{3\,N}
  \!+4\,\left(2\,e^{-4\,\tau}\right)^{3\,N}
  \!+ .\,.\,.\,
  \!+ n^2\,\left(n\,e^{-(n^2)\,\tau}\right)^{3\,N}
  \!+ .\,.\,.\,
    }
    {
    0 
    +\left(e^{-\tau}\right)^{3\,N}
    +\left(2\,e^{-4\,\tau}\right)^{3\,N}
  \!+ .\,.\,.\,
  \!+\left(n\,e^{-(n^2)\,\tau}\right)^{3\,N}
  \!+ .\,.\,.\,
    } 
 \\
%-------------------------------------------------------
 u & \: = \; (\,3\:k\:T\:\tau\,) 
 \:.\:
 \frac{\displaystyle
    1
  \!+4\,\left(2\,e^{-3\,\tau}\right)^{3\,N}
  \!+ .\,.\,.\,
  \!+ n^2\,\left(n\,e^{-(n^2-1)\,\tau}\right)^{3\,N}
  \!+ .\,.\,.\,
    }
    {
    1
    +\left(2\,e^{-3\,\tau}\right)^{3\,N}
  \!+ .\,.\,.\,
  \!+\left(n\,e^{-(n^2-1)\,\tau}\right)^{3\,N}
  \!+ .\,.\,.\,
    } 
 \\
%-------------------------------------------------------
 u_0 & \: \approx \; (\,3\:k\:T\:\tau\,) 
 \:.\:
 \frac{\displaystyle
    1
    }
    {
    1
    } 
 \;=\; 3\:k\:T\:\tau
 \; = \; 
 \frac{3 \: h^2}
 {4\;\pi\;e^{5/3}\;m\;v^{2/3}}
 \quad .\,.\,.\, 
 \quad \mbox{i.e. (\ref{label_eq_47})} \; .
%-------------------------------------------------------
\end{aligned}
\;\;\;\;\right\} 
\!\!\!\!\!\!\label{label_eq_39bis}
\end{equation}
%====================================
The third line is obtained from the second with the approximation 
$T \ll 1$ and thus $\tau \propto 1/T \gg 1$, 
with moreover $N \gg 1$ leading to
$\left(e^{-(n^2-1)\,\tau}\right)^{3\,N} 
\ll e^{-\tau} \ll 1$ by very far.
Therefore only the first two unity terms can be retained at the numerator and the denominator, leading to $u_0 \approx 3\:k\:T\:\tau$ ... QED.
} % {\it \color{blue} \noindent 

  If expression (\ref{label_eq_39}) for the energy of a monatomic gas appears at least debatable compared to Nernst's expression (\ref{label_eq_45}), on closer inspection it turns out to be completely useless. 
Because its size is discontinuous with respect to $\tau$ and $T$, so that a specific heat cannot be defined at all. 

This can be recognized either by a direct examination of the sums in (\ref{label_eq_39}), 
  or more conveniently by considering that the energy of the gas changes 
  in steps with the temperature
not only in a microscopic but also in a macroscopic sense, 
% with the temperature in steps, 
  since according to (\ref{label_eq_36}) the jumps increase in size with increasing atomic number $n$ (i.e. with increasing temperature). 

The hypothesis introduced at the beginning of §\,4 (that the collisions of atoms take place according to the laws of classical mechanics), must therefore be viewed as unfeasible. 
This insight seems to me to be worth investigating.

% ``\, \,''  (\ref{label_eq_24})  \vspace*{-3mm}
% \bigintssss \bigintsss \bigintss \bigints \bigint 
% \dashuline{\,dashing}  -{\it\color{blue}(change)}
%- - - - - - - - - - - -
% $\,$\footnote{$\:$.}  (\ref{label_eq_thermal_1})
% $\,$\footnote{$\:${\it\color{blue}()}.}

\vspace*{-4mm}
\begin{center}
---------------------------------------------------
\end{center}
\vspace*{-3mm}

%========================================
\section{\underline{Third part. A large number of atoms with mutually incoherent} \\ \underline{degrees of freedom.} (p.665-667)}
%========================================
\label{Section-4}
\vspace*{0mm}
% Dritter Teil.
% Eine große Anzahl von Atomen mit gegenseitig inkohärenten Freiheitsgraden. 
% Third part.
% A large number of atoms with mutually incoherent degrees of freedom.

\noindent §\:7. 
% Nachdem im zweiten Teile dieser Arbeit sich die Unhaltbarkeit der Annahme von lauter kohärenten Freiheitsgraden herausgestellt hat, erscheint nunmehr kein anderer Ausweg übrig, als der, die Atombewegungen als inkohärent vorauszusetzen. Dann sind nur die drei Freiheitsgrade jedes einzelnen Atoms unter sich kohärent, d.h.  $$ f=3  f'=3  f''=3 $$ und an die Stelle der Gleichung (35a) tritt die folgende: $$(49)$$, woraus sich durch Zerlegung für ein einzelnes Atom ergibt:  $$(50)$$, und für ein Differentialgebiet: $$(51)$$.
%- - - - - - - - - - - -
% Google: 
After the second part of this work demonstrated the untenability of the assumption of purely coherent degrees of freedom, there now appears to be no other way out than to assume that the atomic movements are incoherent. 

Therefore, only the three degrees of freedom of each individual atom are coherent, i.e. 
$$ f\;=\;3\;, \quad\quad 
  f'\;=\;3\;, \quad\quad 
 f''\;=\;3\;, \quad\quad .\:.\:. $$ 
and equation (\ref{label_eq_35a}) is replaced by the following: 
\begin{align}
\!\!\!\!\!\!
\bigintsss\!\!\!
\bigintsss\!\!\!
\bigintsss\!\!\!
\bigintsss_{\;\;\:0,\:\:0,\:\:0}^{\:n,\:n',\:n''}
\!\!\!\!\!\!\!\!
.\,.\,.\;\: dx \; dy \; dz \:\:
.\,.\,.\;\: d\xi \; d\eta \; d\zeta \:\:
.\,.\,.
\;=\; N\,! \:
\;(n\:h)^{3} \;\: (n'\:h)^{3} .\,.\,.
\;=\; \left(\frac{N}{e} \right)^N
  (n\:h)^{3} \;\: (n'\:h)^{3} .\,.\,.
\label{label_eq_49} \; ,
\end{align}
which gives by decomposition for a single atom: 
\begin{align}
\!\!\!\!\!\!
\bigintsss\!\!\!
\bigintsss\!\!\!
\bigintsss\!\!\!
\bigintsss\!\!\!
\bigintsss\!\!\!
\bigintsss_{\;g=0}^{\:g=g_n}
\!\!\!
\;\: dx \; dy \; dz \; d\xi \; d\eta \; d\zeta 
\;=\; \frac{N}{e} \; g_n^3
\;=\; \frac{N}{e} \; (n\:h)^{3} 
\label{label_eq_50} \; ,
\end{align}
and for a differential field: % region field: 
\begin{align}
\!\!\!\!\!\!
\bigintsss\!\!\!
\bigintsss\!\!\!
\bigintsss\!\!\!
\bigintsss\!\!\!
\bigintsss\!\!\!
\bigintsss_{\;g}^{\:g+dg}
\!\!\!
\;\: dx \; dy \; dz \; d\xi \; d\eta \; d\zeta 
\;=\; \frac{N}{e} \; d\left(g^3\right) 
\label{label_eq_51} \; ,
\end{align}
%- - - - - - - - - - - -
% DeepL: 
% Now that the second part of this work has shown the untenability of the assumption of only coherent degrees of freedom, there seems to be no other way out than to assume that the atomic motions are incoherent. Then only the three degrees of freedom of each individual atom are coherent among themselves, i.e. $$ f=3 f'=3 f''=3 $$ and the equation (35a) is replaced by the following: $$(49)$$, from which, by decomposition, we obtain for a single atom: $$(50)$$, and for a differential field: $$(51)$$.
%- - - - - - - - - - - -

% Integriert man hier über die Raum- und Impulskoordinaten, indem man mit $q$ die Größe der Geschwindigkeit bezeichnet, so ergibt sich: $$AA$$ oder: $$(52)$$.
If one integrates over the space and momentum coordinates by denoting the size of the velocity with $q$, the result is: \vspace*{-2mm}
$$ V \:.\: m^3 \:.\: 
d\left( \frac{4}{3}\:\pi\:q^3 \right)  
\;=\; \frac{N}{e} \; d\left(g^3\right)  $$ 
or {\it\color{blue}(by integration)}:
\vspace*{-3mm} 
\begin{align}
\hspace*{25mm}
q \;=\; \frac{g}{m}
 \; \left( \frac{3}{4\:\pi\:e\:v} \right)^{1/3} \; ,  
 \quad{\it\color{blue}
 \mbox{\it(thus with the constant of integration $=0$?)}}
\label{label_eq_52}
\end{align}
%- - - - - - - - - - - -
%Ebenso: $$BB$$, usw.
and likewise: \vspace*{-3mm}
\begin{align}
q' \;=\; \frac{g'}{m}
 \; \left( \frac{3}{4\:\pi\:e\:v} \right)^{1/3}  
\nonumber
% \label{label_eq_52} 
\; , \quad \mbox{etc.}
\end{align}

%- - - - - - - - - - - -
% Die weitere Berechnung kann wieder ganz auf dem früher eingeschlagenen Wege erfolgen. Zunächst ergibt sich nach (8) für die mittlere Energie im Elementargebiete $(n n' n'')$:  (53), wobei nach (7): (54) und nach (35) und (52):  (55); also durch Ausführung der Integration, mit Berücksichtigung der Grenzen (4): (56).
%- - - - - - - - - - - -
% Google:
% The further calculation can be carried out in the same way as before. First, according to (8) the average energy in the elementary domain $(n n' n'')$ is: (53), where according to (7): (54) and according to (35) and (52): (55); i.e. by carrying out the integration, taking into account the limits (4): (56).
%- - - - - - - - - - - -
% DeepL:
% The rest of the calculation can be done in the same way as before. Firstly, according to (8), the mean energy in the elementary regions is $(n n' n'')$:  (53), where according to (7): (54) and according to (35) and (52):  (55); thus by carrying out the integration, taking into account the limits (4): (56).
%- - - - - - - - - - - -
The further calculation can be carried out in the same way as before. First, according to (\ref{label_eq_8}) the average energy in the elementary domain $(n \:n' \:n'' \:.\,.\,.)$ is: 
\begin{align}
\overline{U}_{\,n\:n'\:n''\:.\,.\,.}
 & \: = \; 
\frac{\displaystyle
\bigintsss_{\,n}^{\,n+1}
\bigintsss_{\,n'}^{\,n'+1}
\bigintsss_{\,n''}^{\,n''+1}
\!\!\!.\:.\:.\;\; U \:\:
d\left(    g^{\,3}\right) 
d\left(  g'^{\,3}\right) 
d\left(g''^{\,3}\right)
\;.\:.\:.\;\;  
}{p\;.\;h^{\,3\,N}} 
\label{label_eq_53} \; ,
\end{align}
where according to (\ref{label_eq_7})
{\color{blue}with $f\,=f'\,=\,f''\,=\,...\,=\,3\,$}: 
\begin{align}
 p
 & \: = \; 
 \left[\, \left( n\,+\,1 \right)^{\,3} 
    \:-\: \left( n \right)^{\,3} 
 \,\right]
 \;.\;
 \left[\, \left( {n'}\,+\,1 \right)^{\,3} 
    \:-\: \left( {n'} \right)^{\,3} 
 \,\right]
 \;.\;
 \left[\, \left( {n''}\,+\,1 \right)^{\,3} 
    \:-\: \left( {n''} \right)^{\,3} 
 \,\right]
 \;.\:.\:.
\label{label_eq_54} \; 
\end{align}
and according to (\ref{label_eq_35}) and (\ref{label_eq_52}): 
\vspace*{-3mm}
\begin{align}
 U & \: = \; 
 \frac{1}{2\:m} \:\:.\: 
    \left( 
       \frac{3}{4\:\pi\:e\:v} 
    \right)^{2/3} \,.\:\;
 \sum_{k=1}^{N}\:
 \left(\:
   g^2 \:+\: {g'}^{\,2} \:+\:{g''}^{\,2} \:+\: .\:.\:.
 \:\right)
\label{label_eq_55} \; ;
\end{align}
i.e. by carrying out the integration, taking into account the limits (\ref{label_eq_4})
{\color{blue}\:with $g\,=n\,h$, $g'\,=n'\,h$, ...\;}: 
\begin{align}
\!\!\!\!
\overline{U}_{\,n\:n'\:n''\:.\,.\,.}
 & \: = \; 
 \frac{3\:h^2}{10\:m} \:\:.\: 
    \left( 
       \frac{3}{4\:\pi\:e\:v} 
    \right)^{2/3} \,.\:\;
 \sum_{k=1}^{N}\:
 \left(\:
   \frac{\left( n\,+\,1 \right)^{\,5}
    \:-\: \left( n \right)^{\,5} }
        {\left( n\,+\,1 \right)^{\,3}
    \:-\: \left( n \right)^{\,3}} 
   \:+\: 
   \frac{\left( n'\,+\,1 \right)^{\,5}
    \:-\: \left( n' \right)^{\,5} }
        {\left( n'\,+\,1 \right)^{\,3}
    \:-\: \left( n' \right)^{\,3}} 
   \:+\: .\:.\:.
 \:\right)
\label{label_eq_56} \; .
\end{align}

{\color{blue} This relationship (\ref{label_eq_56})
is indeed a consequence of the multiple integrations in  (\ref{label_eq_53}), like in the following one with $g\,=n\,h$, with $U$ given by the sum in (\ref{label_eq_55}), with $p$ given by (\ref{label_eq_54}) and $h^3$ (for one of the $N$ atoms) at the denominator:
\begin{align}
\frac{1}{2\:m} 
\bigintsss_{\,n}^{\,n+1} \!\!\!\!
\frac{g^2 \; d\left(g^{\,3}\right)\:/\:h^{\,3}}
{\left( n\,+\,1 \right)^{\,3}
  \:-\:\left( n \right)^{\,3}} 
    \;=
\frac{1}{2\:m} 
\bigintsss_{\,n}^{\,n+1} \!\!\!\!
\frac{3 \; h^5 \; n^4 \; dn\:/\:h^{\,3}}
{\left( n\,+\,1 \right)^{\,3}
  \:-\:\left( n \right)^{\,3}} 
    \;=\;
\left(\frac{3 \; h^2}{10\:m}\right) 
\:.\: 
\left[\:\frac
{\left( n\,+\,1 \right)^{\,5}
  \:-\:\left( n \right)^{\,5}} 
{\left( n\,+\,1 \right)^{\,3}
  \:-\:\left( n \right)^{\,3}} 
\:\right]
\nonumber \; .
\end{align}
}

%- - - - - - - - - - - -
% Dies in (1) eingesetzt, liefert, nach gehöriger reduktion, den Ausdruck der thermodynamischen charakteristischen Funktion, auf ein Atom bezogen: (57), wobei zur Abkürzung gesetzt ist:  (58).
%- - - - - - - - - - - -
% DeepL:
%This, inserted into (1), yields, after appropriate reduction, the expression of the thermodynamic characteristic function, related to an atom: (57), where (58) is set as the abbreviation.
%- - - - - - - - - - - -
% Google:
% Inserting this into (1), after due reduction, gives the expression of the thermodynamic characteristic function related to an atom: (57), whereby the abbreviation is: (58).
%- - - - - - - - - - - -
\noindent
Inserting this {\color{blue}relationship (\ref{label_eq_56})} into (\ref{label_eq_1}), and after due reduction, gives the expression of the thermodynamic characteristic function related to an atom: 
\begin{align}
\psi & \: = \; \frac{\Psi}{N} \; = \; 
k \; \ln\left\{
\;\;\sum_{n=0}^{\infty}
\;\;
 \left[\:
 {\left( n\,+\,1 \right)^{\,3}
  \:-\: \left( n \right)^{\,3}}  
 \:\right]
\:.\:
\exp\left[\:
 -\:\frac{\left( n\,+\,1 \right)^{\,5}
    \:-\: \left( n \right)^{\,5} }
        {\left( n\,+\,1 \right)^{\,3}
    \:-\: \left( n \right)^{\,3}} 
 \:\:.\:\: \sigma
\:\;\right]
\:\;\right\}
\label{label_eq_57} \: ,
\end{align}
where \vspace*{-3mm}
\begin{align}
 \sigma & \: = \; 
 \frac{3\:h^2}{10\:m\:k\:T} \:\:.\: 
    \left( 
       \frac{3}{4\:\pi\:e\:v} 
    \right)^{2/3} 
\label{label_eq_58} \; 
\end{align}
is set as an abbreviation. 

%- - - - - - - - - - - -
% Für hohe Temperaturen ($\sigma \ll 1$) kann man die Summe in (57) als Integral schreiben, nämlich: $$AA$$, und erhält, mit Rücksicht auf (58): $$59$$, was genau mit (42) und dem Tetrodeschen Wert übereinstimmt. 
%- - - - - - - - - - - -
% Google: For high temperatures ($\sigma \ll 1$) you can write the sum in (57) as an integral, namely: $$AA$$, and, taking into account (58): $$59$$, you get what exactly with (42) and the tetrode value corresponds.
%- - - - - - - - - - - -
% DeepL: For high temperatures ($\sigma \ll 1$), the sum in (57) can be written as an integral, namely: $$AA$$, and, taking into account (58): $$59$$, which agrees exactly with (42) and Tetrode's value. 
%- - - - - - - - - - - -
\dashuline{For high temperatures} ($\sigma \ll 1$) you can write the sum in (\ref{label_eq_57})  as an integral, namely: 
\begin{align}
 \psi & \: = \; k \:.\:
  \ln\left[
    \;\;
    \bigintsss_{\;\;0}^{\:\infty}
    \;\left[\:\:
       3 \;\: n^2 \:\: 
       \exp\left(\:-\,\frac{5}{3}\:n^2\:\sigma\:\right)
    \:\right]
    \:.\: dn
\:\;\right]
\nonumber % \label{label_eq_41} 
\; , 
\end{align}
and taking into account (\ref{label_eq_58}): 
\begin{align}
 \psi & \: = \; 
   \frac{3}{2} \:\: k
   \; \ln\left( \, 
   \frac{2\:\pi\:m\:k\:T \:(e\:v)^{2/3}}{h^2}
   \, \right) 
\label{label_eq_59} 
\; ,
\end{align}
which agrees exactly with (\ref{label_eq_42}) and Tetrode's value{\it\color{blue}, and thus with the same absolute entropy (\ref{label_eq_44})\,:
\vspace*{-1mm}
\begin{align}\tag{44}
\boxed{\:
 s \; = \; 
\frac{3}{2}\; k \;
\ln\left(\:
\frac{2\:\pi\:e^{5/3}\:m\:v^{2/3}\:k\:T}{h^2}
\:\right)
\:} 
\label{label_eq_44_bis} 
\; .
\end{align}
}
\vspace*{-2mm}

%- - - - - - - - - - - -
% Für tiefe Temperaturen ($\sigma \gg 1$) dagegen ergibt sich aus (57): $$AA$$, also nach (2) die Nullpunktsenergie:  $$60$$, gegen welche der Nernstsche Wert (48) im Verhältnis $$BB$$ größer ist. 
%- - - - - - - - - - - -
% DeepL: For low temperatures ($\sigma \gg 1$), on the other hand, (57) yields: $$AA$$, i.e. according to (2) the zero point energy: $$60$$, against which the Nernst value (48) is greater in the ratio $$BB$$. 
%- - - - - - - - - - - -
% Google: For low temperatures ($\sigma \gg 1$), on the other hand, from (57): $$AA$$, so according to (2) the zero point energy: $$60$$, against which the Nernst value (48) in the ratio $ $BB$$ is larger.
%- - - - - - - - - - - -
\dashuline{For low temperatures} ($\sigma \gg 1$), 
on the other hand, (\ref{label_eq_57}) 
{\color{blue}with only the first-order term $n=0$}
yields: 
$$ 
 \psi 
 {\color{blue}\,
 \;\: \approx \;\:  
 k\:\ln\left\{\:\exp\,[\:-\:\sigma\:]\:\right\}
 \:}
 \; = \; 
 -\:k\;\sigma 
 {\color{blue}\,
 \;\: = \;\:  
 -\:\frac{3\:h^2}{10\:m} \:\:.\:\: 
    \frac{1}{T} \:\:.\: 
    \left( 
       \frac{3}{4\:\pi\:e\:v} 
    \right)^{2/3} 
 }
 \; ,
$$ 
i.e., according to (\ref{label_eq_2}), 
{\bf\dashuline{the zero point energy}}: 
\vspace*{-3mm}
\begin{align}
\boxed{\;
u_0 \;=\; T^2\:.\:\,\frac{\partial\:\psi}{\partial\:T}
\;=\:
 \frac{3\:h^2}{10\:m} \:\:.\: 
    \left( 
       \frac{3}{4\:\pi\:e\:v} 
    \right)^{2/3} 
\;}
{\color{blue}\;\;=\;
 \overbrace{
 \left(\,
 \frac{3 \: h^2}
 {4\;\pi\;e^{5/3}\;m\;v^{2/3}}
 \,\right)}^{\displaystyle \mbox{Eq.~(\ref{label_eq_47})}}
 \:.\:
 \overbrace{
 \left[\,
 \frac{2\;\pi\;e}{5}
 \left(
 \frac{3}{4\;\pi}
 \right)^{2/3}
 \,\right]}^{\displaystyle \approx 1.315}
}
\label{label_eq_60} \: 
\end{align}
against which the Nernst value (\ref{label_eq_48}) is greater in the ratio 
\vspace*{-1mm}
\begin{align}
    \frac{5}{4\:\pi} 
    \:\:.\: 
    \left( 
       \frac{4\:\pi\:e}{3}
    \right)^{2/3} 
 \; \approx \; 
 2.014
\nonumber % \label{label_eq_58} 
\; .
\end{align}

\vspace*{-4mm}
\begin{center}
---------------------------------------------------
\end{center}
\vspace*{-3mm}

{\it\color{blue}\dashuline{Notes of P. Marquet}: an additional comparison with (\ref{label_eq_47}) shows that (\ref{label_eq_60}) is greater in the ratio $1.315$. 
Moreover, it is possible to write (\ref{label_eq_60}) in terms of either the $2$ (molar-)variables $M=N\:m$ and $V=N\:v$:
\vspace*{-3mm}
\begin{align}\tag{60-bis}
\boxed{\;
U_0 \;=\; N\:u_0 \;=\; 
\left[\:
 \frac{3\:h^2}{10} \:\:.\:\: 
  N^{8/3} \:\:.\: 
    \left( 
       \frac{3}{4\:\pi\:e} 
    \right)^{2/3}  
 \:\right] \:.\:\: 
 \frac{1}{M\;V^{2/3}}
 \; \approx \;
 \frac{\left[\: 6.73\:10^{\,-5}\:\right]}{M\;V^{2/3}}
 \;\mbox{~J~kg${}^{\,-1}$}
 \; , 
\;}
\label{label_eq_60_bis} \: 
\end{align}
or equivalently in terms of the $3$ variables $T$, $M=N\:m$ and 
$P = R \: T / V = N \: k \: T / V$:
\vspace*{0mm}
\begin{align}\tag{60-ter}
\boxed{\;
U_0 \;=\; N\:u_0 \;=\; 
\left[\:
 \frac{3\:h^2}{10} \:\:.\:\: 
 \frac{N^2}{k^{2/3}} \:\:.\: 
    \left( 
       \frac{3}{4\:\pi\:e} 
    \right)^{2/3}  
 \:\right] \:.\:\: 
 \frac{P^{2/3}}{M\;T^{2/3}}
 \; \approx \;
 \frac{\left[\: 1.64\:10^{\,-5} \:\right]\,P^{2/3}}{M\;T^{2/3}}
 \;\mbox{~J~kg${}^{\,-1}$}
 \; . 
\;}
\label{label_eq_60_ter} \: 
\end{align}
The two numerical values 
$6.730\,494\,791\:10^{\,-5}
\approx 6.73\:10^{\,-5}$ 
in (\ref{label_eq_60_bis}) and 
$1.639\,926\:10^{\,-5}
\approx 1.64\:10^{\,-5}$ 
in (\ref{label_eq_60_ter}) are computed from the exact values for $h$, $k$ and $N$ set in May 2019 for five of the seven SI base units specified in the 
%- - - - - - - - - - - - - - - - - - - - - - - - - - -
International System of Quantities$\,$\footnote{$\:${\it\color{blue}See:
[\,\url{https://en.wikipedia.org/wiki/2019_redefinition_of_the_SI_base_units}\,] 
and
[\,\url{https://www.bipm.org/en/publications/si-brochure/}\,]}.}
%- - - - - - - - - - - - - - - - - - - - - - - - - - -
(along with $c$, already defined as an exact value): 
\vspace*{1mm} \\ \hspace*{5mm}$\;\;\;\bullet$
the Planck constant 
$(b=h=6.626\,070\,15 \times 10^{-34}$~J~s$\,)$; 
\vspace*{1mm} \\ \hspace*{5mm}$\;\;\;\bullet$
the Planck\,(--Boltzmann\,) constant 
$(b/a=k=1.380\,649 \times 10^{-23}$~J~K${}^{\,-1}\,)$;    
\vspace*{1mm} \\ \hspace*{5mm}$\;\;\;\bullet$
the Avogadro constant 
$(N=6.022\,140\,76\times 10^{23}$~mol${}^{\,-1}\,)$;
\vspace*{1mm} \\ \hspace*{5mm}$\;\;\;\bullet$
the elementary electric charge 
$(q_e=1.602\,176\,634 \times  10^{-19}$~C$\,)$; and
\vspace*{1mm} \\ \hspace*{5mm}$\;\;\;\bullet$
the (previously exact) velocity of light in vacuum
$(c=2.997\,924\,58 \times 10^{8}$~m~s${}^{\,-1}\,)$.
\vspace*{1mm} \\
%leading to
%\underline{$R_{\ast} = 8.314\,462\,62$~J~K${}^{-1}$~mol${}^{-1}$},
%\underline{$\sigma   = 5.670\,374\,42 \times 10^{-8}$},
%\underline{$c_2        = 1.438\,776\,88$~cm~K}, 
%\underline{$K_{\rm ST} = 18.222\,850\,85$},
%\underline{$C_R        = 40.275\,324\,09 
%          \times 10^{-47}$~J~s${}^{2}$~K${}^{2}$},
The accuracy is of $\:9$ digits for $h$ and $N$ --and thus for the constant in (\ref{label_eq_60_bis}).
The accuracy is of $\:7$ digits only for $k$ --and thus for the constant in (\ref{label_eq_60_ter}).

Note that the energies at $0$~K in units of J~mol${}^{\,-1}$ can be computed from (\ref{label_eq_60_bis}) or (\ref{label_eq_60_ter}) via the relationship ``\,$M\:U_0$'' 
equal to 
$[\: 6.73\:10^{\,-5}\:]\,/\,V^{2/3}$ and
$[\: 1.64\:10^{\,-5} \:]\,.\,(P/T)^{2/3}$, respectively.

}

\vspace*{0mm}
\begin{center}
---------------------------------------------------
\end{center}
\vspace*{-3mm}

{\color{blue}\it(Conclusion of Max Planck:)}

For arbitrary temperatures, the execution of the summation in (\ref{label_eq_57}) causes at first a certain difficulty, but this can soon be overcome, and I hope to soon be able to communicate a series expansion convenient for the calculation of both the specific heat and 
the pressure{\it\color{blue}$\,$\footnote{$\:${\it\color{blue}I have not found the paper where these computations of Max Planck may have been published after 1916 / P. Marquet)\,}}}.

As far as I can see, the resulting formulas form the only possible result of applying the quantum hypothesis to the thermodynamic behavior of a monatomic gas, the density of which is so low that no other form of energy can be considered apart from the kinetic energy of the atoms.

% ``\, \,''  (\ref{label_eq_24})  \vspace*{-3mm}
% \bigintssss \bigintsss \bigintss \bigints \bigint 
% \dashuline{\,dashing}  -{\it\color{blue}(change)}
%- - - - - - - - - - - -
% $\,$\footnote{$\:$.}  (\ref{label_eq_thermal_1})
% $\,$\footnote{$\:${\it\color{blue}()}.}

\vspace*{-4mm}
\begin{center}
---------------------------------------------------
\end{center}
\vspace*{-3mm}

%==============================
\end{document}